\documentclass{aa}  

\usepackage{graphicx}
\usepackage{txfonts}
\usepackage{xcolor}
\newcommand{\xmm}{\textit{XMM-Newton}}
\begin{document}

   \title{Black hole mass of central galaxies and cluster mass correlation in cosmological hydro-dynamical simulations}
 
   \author{L. Bassini\inst{1,2,3,4},
          E. Rasia\inst{2,3},
          S. Borgani\inst{1,2,3,4},
          C. Ragone-Figueroa\inst{2,5,6},
          V. Biffi\inst{1,2,7},
          K. Dolag\inst{8,9},
          M. Gaspari\inst{10}\thanks{\textit{Lyman Spitzer Jr.} Fellow},
          G. L. Granato\inst{2,5,6},
          G. Murante\inst{2},
          G. Taffoni\inst{2},
          L. Tornatore\inst{2}
          }

   \institute{Dipartimento di Fisica dell’Università di Trieste, Sezione di Astronomia, via Tiepolo 11, I-34131 Trieste, Italy
         \and
           INAF, Osservatorio Astronomico di Trieste, via Tiepolo 11, I-34131, Trieste, Italy
        \and
                    IFPU - Institute for Fundamental Physics of the Universe, Via Beirut 2, 34014, Trieste, Italy
        \and  
        INFN, Instituto Nazionale di Fisica Nucleare, Trieste, Italy
        \and
        Universidad Nacional de C\'{o}rdoba. Observatorio Astron\'{o}mico de C\'{o}rdoba. C\'{o}rdoba, Argentina
        \and
         CONICET. Instituto de Astronomía Te\'{o}rica y Experimental. C\'{o}rdoba, Argentina
        \and
        Harvard-Smithsonian Center for Astrophysics, 60 Garden St., Cambridge, MA 02138, USA
        \and 
            University Observatory Munich, Scheinerstraße 1, D-81679 Munich, Germany
        \and
            Max-Planck-Institut f{\"u}r Astrophysik, Karl-Schwarzschild-Stra{\ss}e 1, 85748 Garching bei M{\"u}nchen, Germany
        \and 
            Department of Astrophysical Sciences, Princeton University, 4 Ivy Lane, Princeton, NJ 08544-1001, USA.
              }
            
 \date{Received XXX; accepted XXX}
 \titlerunning{$M_{\rm BH}-M_{500}$ correlation in simulated clusters}
 \authorrunning{Bassini et al.}

 
  \abstract
   {The correlations between the properties of the brightest cluster galaxy (BCG) and the mass of its central super-massive black hole (SMBH) have been extensively studied from a theoretical and observational angle. More recently, relations connecting the SMBH mass and global properties of the hosting cluster, such as temperature and mass, were observed.}
   {We investigate the correlation between SMBH mass and cluster mass and temperature, their establishment and evolution. We compare their scatter to that of the classical $M_{\rm BH}-M_{\rm BCG}$ relation. Moreover, we study how gas accretion and BH-BH mergers contribute to SMBH growth across cosmic time.}
   {We employed 135 groups and clusters with a mass range $1.4\times 10^{13}M_{\odot}-2.5\times 10^{15} M_{\odot}$ extracted from a set of 29 zoom-in cosmological hydro-dynamical simulations where the baryonic physics is treated with various sub-grid models, including feedback by active galactic nuclei.}
   {In our simulations we find that $M_{\rm BH}$ correlates well with $M_{500}$ and $T_{500}$, with the scatter around these relations compatible within $2\sigma$ with the scatter around $M_{\rm BH}-M_{\rm BCG}$ at $z=0$. The $M_{\rm BH}-M_{500}$ relation evolves with time, becoming shallower at lower redshift as a direct consequence of hierarchical structure formation. On average, in our simulations the contribution of gas accretion to the total SMBH mass  dominates for the majority of the cosmic time ($z>0.4$), while in the last 2 Gyr the BH-BH mergers become a larger contributor. During this last process, substructures hosting SMBHs are disrupted in the merger process with the BCG and the unbound stars enrich the diffuse stellar component rather than increase BCG mass. }
   {From the results obtained in our simulations with simple sub-grid models we conclude that the scatter around the $M_{\rm BH}-T_{500}$ relation is comparable to the scatter around the $M_{\rm BH}-M_{\rm BCG}$ relation and that, given the observational difficulties related to the estimation of the BCG mass, clusters temperature and mass can be a useful proxy for the SMBHs mass, especially at high redshift.}

   \keywords{galaxy: clusters: intracluster medium --
               galaxy: nuclei --
               method: numerical --
               X-ray: galaxies: cluster
               }

   \maketitle
%
\section{Introduction}

It is well known that galaxies of every morphology host super massive black holes (SMBHs) at their center. Interestingly the mass of these  SMBHs correlate well with a number of bulge properties of the host galaxy such as bulge stellar mass (e.g., \citealt{MAGORRIAN1998}; \citealt{2003MARCONI}; \citealt{2004HARING}; \citealt{2009HU}; \citealt{2011SANI}; \citealt{2013KORMENDY} for a review), 
bulge luminosity (e.g., \citealt{1995KORMENDY}; \citealt{2002MCLURE}; \citealt{2009HU}; \citealt{2011SANI}), and {especially the bulge stellar velocity dispersion} (e.g., \citealt{2000FERRARESE}; \citealt{2000GEBHARDT}; \citealt{2001MERRITT}; \citealt{2002TREMAINE}; \citealt{2006WYITHE}; \citealt{2008HU}; \citealt{2009GULTEKIN}; \citealt{2012BEIFIORI}; \citealt{2013mcconnel}). These correlations are as tight as to be often used to estimate the SMBH mass when dynamical measurements are not available and  suggest a co-evolution between SMBH and the hosting galaxy, although the main physical processes involved are still debated. In the last 20 years many possibilities have been proposed. The most commonly suggested and widely accepted mechanism is the active-galactic-nuclei (AGN) feedback. In this scenario gas settles around the SMBH radiating energy at a rate of $\sim \eta \dot{M}c^2$, with $\eta \sim 0.1$. If the feedback is strong enough to overcome the binding energy, cold gas is expelled from the galaxy halting both star formation and accretion around the central SMBH (e.g., \citealt{1999FABIAN}; \citealt{2004GRANATO}; \citealt{2005DIMATTEO}; \citealt{2006HOPKINS}). However, other authors have shown that the M$_{\rm BH}$-M$_{\star}$ relation can arise or be contributed by non-causal processes. In this scenario the observed M$_{\rm BH}$-M$_{\star}$ relation follows from hierarchical galaxy mergers starting from an uncorrelated distribution of M$_{\rm BH}$ and M$_{\star}$ (e.g., \citealt{2007PENG}; \citealt{2011KUND}).

For massive elliptical galaxies, it has also been suggested that the main correlation is not with the bulge properties but with the host dark matter halo (e.g., \citealt{2002FERRARESE}; \citealt{2009BANDARA}; \citealt{2010BOOTH}; \citealt{2011MARTA}). 
In this respect, between all galaxies a special position is occupied by the brightest cluster galaxies (BCGs), which sit at the center of galaxy clusters and host the most massive SMBHs in the universe. Since the luminosity or mass of these galaxies have been found to relate to the hosting halo both in observational (e.g, \citealt{2004LIN}; \citealt{2008BROUGH}) and numerical (e.g., \citealt{2018CINTHIA}) studies, a correlation between $M_{\rm BH}$ and cluster mass is expected (e.g., \citealt{2009MITTAL}), although the SMBH and cluster global properties ought not be necessarily connected. The argument is not completely settled as observations of another class of objects, bulgeless spiral galaxies, suggest that SMBHs do not correlate directly with dark matter, the latter being estimated from the circular rotation velocities of gas in the outskirts (e.g., \citealt{2011NATURE}; \citealt{2015SABRA}). 

\cite{2005CHURAZOV} proposed a toy model for the evolution of massive elliptical galaxies and their central SMBHs. In this simple model, the equivalence between gas cooling and AGN heating leads to a correlation between SMBH mass and the stellar velocity dispersion (assuming a rough $\sigma_{\rm vel}^2 \sim T$, we can further convert it into a temperature correlation).
 \cite{GS17} proposed that the feeding via chaotic cold accretion (CCA) boosts the AGN feedback in a tight self-regulated feedback loop that prevents catastrophic cooling flows for several gigayears, driving a direct physical connection between the SMBH mass and intra cluster medium (ICM) properties, such as plasma temperature and luminosity. With a sample of groups and clusters of galaxies, \cite{2018BOGDAN} corroborated this prediction, further finding a scatter in the $M_{\rm BH}-T_{500}$ relation that is lower than that of the $M_{\rm BH}-M_{\star}$ relation.
 \cite{2019PHIPPS} enlarged the sample of \cite{2018BOGDAN}, focusing on the cluster scales and retrieving the SMBH mass from the AGN fundamental plane (\citealt{2003MERLONI}). \cite{2019GASPARI} based their sample on SMBHs with direct dynamical mass measurements, considering different type of galaxies (BCGs, BGGs, ETGs, S0s, and massive LTGs) and exploring correlations with thermodynamic properties  in addition to the X-ray luminosity and temperature (such as the gas pressure, gas mass, and the parameter $Y_X=M_{gas}\times T$) over different extraction radii (galactic, core, $R_{500}$).  
%
Finally, \cite{2019MNRAS.tmpL.117L} considered a sample of 41 ETGs and confirmed the tight correlation between the SMBH mass and the X-ray temperature measured within the effective radius.



All these recent works suggest that the growth of SMBHs in BCG 
is regulated by physical processes that also influence the thermo-dynamical properties of the ICM. Various phenomena dominate at different scales, from few kiloparsec to megaparsec. In this work we focus on the largest scale. 
Specifically, in this paper, we investigate the correlation between the mass of SMBHs in BCGs and the global properties of the hosting cluster, such as temperature and mass measured within $R_{500}$ (defined as the radius of the sphere of mass $M_{500}$ whose  enclosed density is $500$ times the critical density of the Universe at a specific redshift, $3H(z)^2$/$8\pi G$)
by employing a set of cosmological hydro-dynamical simulations centered on massive clusters. These zoom-in simulations include a number of sub-grid models for radiative cooling, star formation and associated feedback, metal enrichment, and chemical evolution and they implement recipes for SMBH accretion and consequent AGN feedback (\citealt{2013RAGONE}). In past works with a similar set of simulations, we showed that the AGN feedback at the center of galaxy clusters leads to an appropriate description of the observed ICM thermodynamic quantities (such as entropy, gas density, temperature, and thermal pressure) and metallicity (\citealt{2015RASIA}; \citealt{2017PLANELLES}; \citealt{2017BIFFI}). Given these previous results, we investigate further these simulated regions to study the physical processes that tie SMBH to the hosting cluster. In particular in this work we aim at answering the following questions: 1) Do numerical simulations reproduce the observed $T_{\rm 500}-M_{\rm BH}$ and $M_{\rm 500}$-$M_{\rm BH}$ relations? 2) Which are the processes that lead to the observed relations? 3) Do the relations evolve with redshift? 4) Through which channels (e.g., gas accretion or BH-BH mergers) do SMBHs grow in time? 5) Is $M_{\rm 500}$ as appropriate as $M_{\rm BCG}$ to probe $M_{\rm BH}$?

The paper is structured as follows: in Sect. 2 we describe the numerical simulations employed.  In Sect. 3 we detail how the quantities of interest are computed from simulations and the method employed for linear fitting. In Sect. 4 we present our results, that we discuss in Sect. 5 before concluding in Sect.~6.

\section{Simulations}\label{section:simulations}

Our analysis is based on a set of 29 cosmological and hydro-dynamical zoom-in simulations centered on massive galaxy clusters evolved in a $\Lambda$CDM model with parameters: $\Omega_{\rm m}=0.24$, $\Omega_{\rm b}=0.04$, $n_{\rm s}=0.96$, $\sigma_8=0.8$, and $H_0=100$ $h$ km s$^{-1}$ Mpc$^{-1}=72$ km s$^{-1}$ Mpc$^{-1}$. These regions were selected from a parent $N$-Body cosmological volume of 1 $h^{-3}$ Gpc$^3$ and re-simulated at higher resolution with the inclusion of baryons (for a detailed description of the initial conditions see \citealt{Bonafede}). 

The re-simulated regions are centered around the 24 most massive clusters of the parental box with mass $M_{200}$ $ \geq 8 \times 10^{14}h^{-1}$ M$_{\odot}$ and 5 isolated groups with $M_{200}$ within $[1 - 4] \times 10^{14} h^{-1}$ M$_{\odot}$. In the high-resolution regions the mass of DM particles is $m_{\rm DM}=8.47 \times 10^8$ $h^{-1}$ M$_{\odot}$ and the initial mass of the gas particle is $m_{\rm gas}=1.53 \times 10^8$ $h^{-1}$ M$_{\odot}$. The Plummer equivalent gravitational softening for DM particles is set to $\epsilon = 5.6 h^{-1} {\rm kpc}$ in comoving units at redshift higher than $z=2$ and in physical units afterward. The gravitational softening lengths of gas, stars, and black hole particles are fixed in comoving coordinates to $5.6$ $h^{-1}$ kpc, $3$ $h^{-1}$ kpc, and $3$ $h^{-1}$ kpc, respectively.

The simulations were carried out with the code GADGET-3, a modified version of the Tree-PM Smoothed-Particle-Hydrodynamics (SPH) public code GADGET2 (\citealt{GADGET2005}). Our simulations are performed with an improved version that accounts for modifications of the hydrodynamic scheme to better capture hydro-dynamical instabilities (see \citealt{BECK2016}). These changes include a higher order kernel function, a time dependent artificial viscosity scheme, and a time dependent artificial conduction scheme.

The set of zoom-in simulations treats unresolved baryonic physics through various sub-grid models. A detailed description can be found in \cite{2017PLANELLES} or \cite{2017BIFFI}; we briefly summarize here the main aspects. The prescription of metal-dependent radiative cooling follows \cite{WIERSMA2009}. The model of star formation and associated feedback prescriptions are implemented according to the original model by \cite{SPRINGEL2003}, and metal enrichment and chemical evolution following the formulation by \cite{LUCA2007}. The yields used in our simulations are specified in \cite{2018BIFFI}. The AGN feedback model is implemented as described in Appendix A of \cite{2013RAGONE} with one important modification \citep[see,][]{2018CINTHIA}: the distinction between cold mode and hot mode gas accretion \citep[see also][]{2015RASIA}. In practice, the gas accretion is the minimum between the Eddington limit and the $\alpha$-modified Bondi accretion rate:
\begin{equation}
	\dot{M}_{{\rm Bondi,} \alpha} = \alpha \frac{4 \pi G^2 M_{\rm BH}^2 \rho}{(c_{\rm s}^2 + \nu_{\rm BH}^2)^{3/2}} ,
\label{eq:bondi}
\end{equation}
with $\alpha$ equal to 10 and 100 for hot ($T>5\times 10^5$ K) and cold ($T<5\times 10^5$ K) gas respectively. The $100\times$ boost during the cold gas accretion mimics the impact of CCA (e.g., \citealt{GS17}). $M_{\rm BH}$ is the SMBH mass. All other quantities relate to the gas and are smoothed over 200 gas particles
with a kernel centered at the position of the black hole: $\rho$ is the gas density, $c_{\rm s}$ is the sound speed of the gas surrounding the SMBH, and $\nu_{\rm BH}$ is the relative velocity between the SMBH and bulk velocity of the gas. We note that For the SMBHs considered in this paper (e.g., SMBHs at the center of BCGs) the radius which enclose 200 gas particles is $\sim 50$ kpc at $z=0$. This radius decreases at higher redshift, having a value of $\sim 30$ kpc at $z=2$. In order to avoid wandering black holes, they are re-positioned  at each time step on the position of the most bound particle within the SMBH softening length. This calculation is restricted to particles with relative velocity with respect to the SMBH below $300$ s$^{-1}$ km. This condition avoids that the SMBH particle "jumps" into a close flyby structure that would displace it from the cluster center.

A fraction $\epsilon_{\rm r}$ of the energy associated to the gas directly fueling the SMBH through accretion is radiated away and a fraction $\epsilon_{\rm f}$ of this energy is thermally coupled to the surrounding gas particles. The value of $\epsilon_{\rm r}$ is fixed to $0.07$, while that of $\epsilon_f$ depends on the mode of the AGN: during the quasar mode, meaning for $\dot{M}_{\rm BH}/\dot{M}_{\rm Edd}>0.01$, $\epsilon_{\rm f}=0.1$, while during the radio mode $\epsilon_{\rm f}$ is increased to $0.7$ \citep[see,][]{2018CINTHIA}. The exact values of both parameters are chosen to reproduce the observed correlation between stellar mass and SMBH mass in galaxies (see Fig. \ref{fig:magorrian}). Recently some studies (e.g., \cite{2016SHANKAR}; \cite{2019SHANKAR}) pointed out that the currently observed relation between stellar mass and SMBH mass may be biased high. We study the effect of artificially increasing feedback parameters on our results in the Appendix.


 SMBHs of mass $4.4 \times 10^5$ M$_{\odot}$ are seeded at the position of the most bound particle of the structures that, identified by the Friends-of-Friends algorithm, simultaneously satisfy all the following conditions: the stellar mass of the structure is greater than $2.2 \times 10^{10} M_{\odot}$ and it is higher than 5 percent of the dark-matter halo mass; the ratio between the gas mass and the stellar mass is higher than $0.1$;  no other central SMBH is already present. The mass of the seeding is consistent with the expectation of the direct collapse. Under these conditions, the seeding of the SMBH happens in galaxies that have enough gas to promptly feed it.
Two SMBH particles merge whenever their relative velocity $v_{\rm rev}$ is smaller than $0.5 \times c_s$ and their distance $r$ is less than twice the SMBH softening length. When a BH-BH merger happens, the SMBH particle of the most massive SMBH gains the mass of the merged SMBH.
The strategy to position the SMBH, the recipe to implement the AGN feedback, and slightly larger softening lengths are the only differences with respect to the simulation set-up of the runs presented in \cite{2015RASIA}, \cite{2017BIFFI}, \cite{2017PLANELLES}, \cite{2018BIFFI}, \cite{2018TRUONG}.

\subsection{Calibration of AGN feedback model}

\begin{figure}
	\includegraphics[width=\columnwidth]{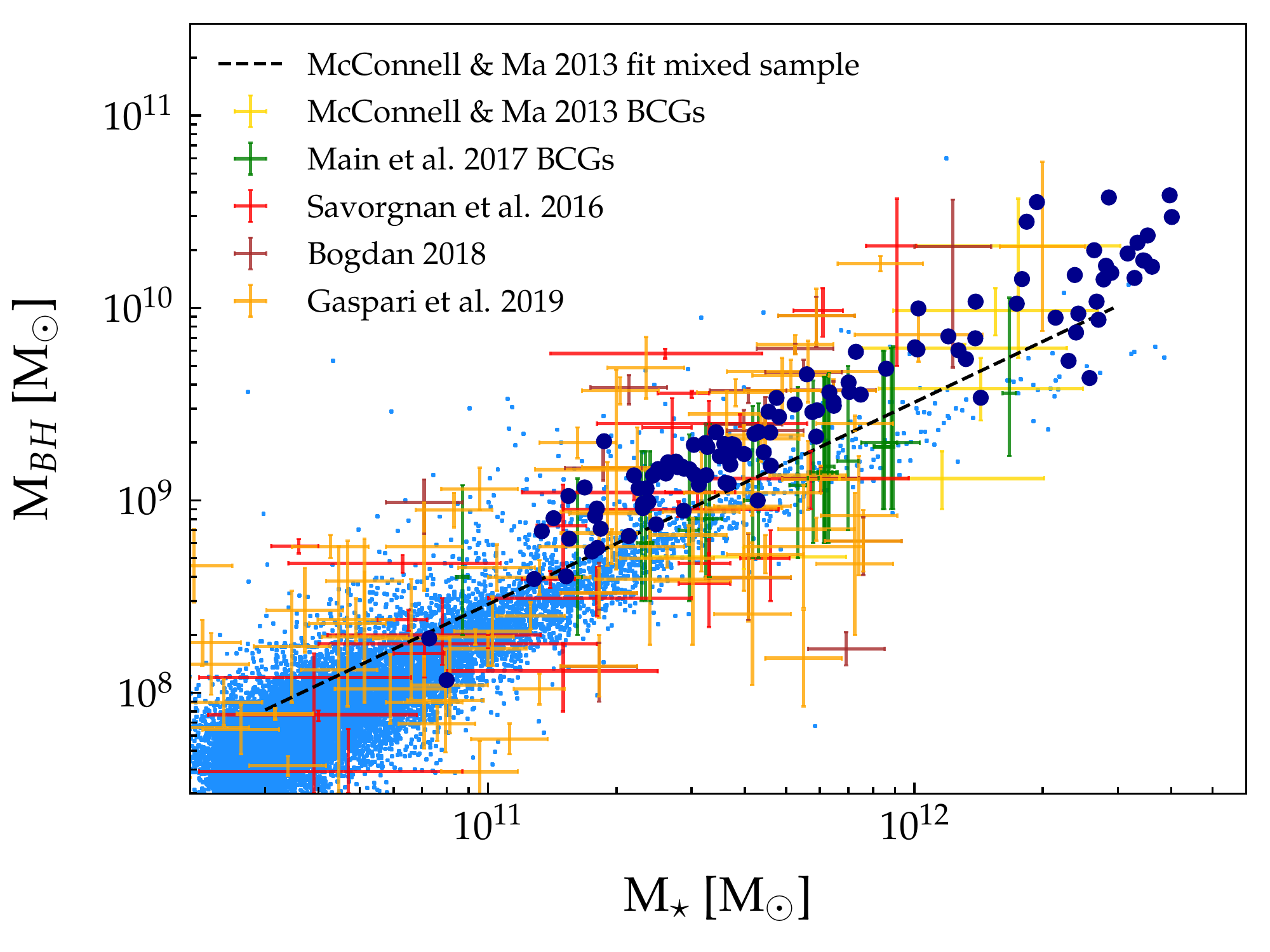}
    \caption{Correlation between stellar mass and SMBH mass in observations and simulations.  Small light-blue points represent non-BCG simulated galaxies, large black dots represent simulated BCGs. Yellow, green, red, brown, and orange crosses represent the observational  data with their error bars taken from \protect\cite{2013mcconnel}, \protect\cite{2017MAIN}, \protect\cite{Savorgnan2016}, \protect\cite{2018BOGDAN}, and \protect\cite{2019GASPARI} respectively. The dashed black line is a linear best-fit of the sample of different type of galaxies by \protect\cite{2013mcconnel}. See text for details about $M_{\rm BH}$ and $M_{\star}$ definition and measurement. }
    \label{fig:magorrian}
\end{figure}

In Fig.~\ref{fig:magorrian} we show the calibration of the AGN feedback model used in the simulations. This is based on the correlation between the stellar mass of galaxies, $M_{\star}$, and the mass of their central SMBHs, $M_{\rm BH}$. In the figure the small light-blue points represent non-central simulated galaxies identified by Subfind (\citealt{Dolag2009}), while dark-blue dots represent simulated BCGs. The stellar masses of the BCGs are defined as the mass enclosed in a sphere of radius $ 0.1 \times R_{500}$ around the position of the central SMBH, while total stellar masses of non-BCGs are given as an output by Subfind. 

To calibrate the parameters for the AGN-feedback model we aimed at reproducing the entire  $M_{\rm BH}-M_{\star}$ relation including the majority of simulated non-BCG galaxies. These are, in particular, compared with the observational data from \cite{2013mcconnel} represented in the figure by the dashed line. 

In the plot we also include other observational data, namely the BCGs from \cite{2013mcconnel} and the samples from \cite{Savorgnan2016}, \cite{2017MAIN}, \cite{2018BOGDAN},  and \cite{2019GASPARI}. In \cite{2017MAIN} SMBH masses are computed from $K$-band luminosities using the relation $\log(M_{\mathrm{BH}}/{\rm M}_{\odot}) = -0.38(\pm 0.06)(M_{\rm K}+24) + 8.26 (\pm 0.11)$ suggested by \cite{2007GRAHAM} and extracted from a sample of elliptical but not BCGs. In all the other works the mass of the SMBHs are derived from dynamical measurements. The BCG masses in \cite{2013mcconnel} and \cite{2018BOGDAN} are part of a compilation from previous literature and  we refer to the original papers and references therein for further information on the methods employed to infer the stellar masses. In \cite{Savorgnan2016} the stellar masses are computed from bulge luminosities assuming a constant mass-to-light ratio,  while in \cite{2019GASPARI} it is assumed a variable $M_{\star}/L_K$ scaling as a function of the stellar velocity dispersion (see \citealt{2019GASPARI} for details). It should be noted that from \cite{Savorgnan2016} we used only ellipticals, that are not necessarily BCGs. In \cite{2017MAIN} the stellar masses are computed from K-band luminosity using the relation log$(M/L_{\rm K}) = -0.206 + 0.135 (B-V)$ given by \cite{2003KBAND}.

Our main condition for deciding on the AGN parameters is that the observed correlation between SMBHs mass and non-BCG galaxies (dashed line) passes through the bulk of the simulated galaxies (small blue points). We also care for an overall agreement at the BCG scales but with less emphasis because of the scatter of the observed sample \citep{2013mcconnel} is high  at the high mass end. Regarding the BCGs, we find that the simulated BCGs are in a good agreement with observational data at both ends of the mass range, but that the  simulated points tend to stay above observational data around $M_{\star} = 10^{12} {\rm M}_{\odot}$, although still inside their error bars. This discrepancy does not necessarily highlight a poor description of the simulations since several factors need to be considered for a proper comparison. First, the SMBH masses are computed by adopting different methods. For example, those extracted by \cite{2017MAIN} are calibrated using a relation that does not include BCGs and, indeed, they are more aligned with the non-BCG sample. Second, BCG stellar masses are computed using different apertures in simulations and in the various observational samples. Furthermore, measurements of stellar mass from different works can disagree due to the different assumptions made during the data analysis, such as the assumed initial mass function, the adopted stellar mass-to-light ratio, distances, and beam aperture. Dynamical measurements can significantly disagree in particular when considering different tracers, such as stars versus circumnuclear gas, including different gas phases as warm and ionized versus cold and molecular gas. The resulting differences among catalogs can be comparable to the separation between simulated and observed data points.
An example is clearly represented in the figure by NGC 4889, the galaxy with the most massive SMBH. This object is present in the \cite{Savorgnan2016}, \cite{2013mcconnel}, and \cite{2018BOGDAN} samples and, while M$_{\rm BH}$ is identical because taken from the same source in the literature, the estimated BCG mass can be different even by a factor of $\sim 2$. This emphasizes the intrinsic difficulty in defining the BCG stellar masses and, at the same time, it quantifies a possible level of stellar mass difference among different works.

\section{Method and samples}\label{methods}

To investigate possible correlations between the central SMBHs and the global cluster properties we need to extract the cluster masses and temperatures from the simulated regions. In addition, we need to calculate the SMBH mass and the two contributions to its growth: the accretion of the surrounding gas and the merger with other SMBHs.  In the following, after specifying the definition of the cluster center, we describe how all these quantities are computed.\\

\noindent {\it Cluster center}. As mentioned in the previous section, the SMBHs are always positioned at the location of the most bound particle that should identify the center of the host halo. Therefore, we followed the SMBHs back in time to identify the position of the hosting halo center. For this goal, we saved at z=0 the unique
identification number of the most massive SMBH particle which is within 10 kpc from the cluster minimum of the potential well, as identified by Subfind. Subsequently, we   tracked it back in time to the epoch of its seeding. At each time, we checked that the SMBH is, as expected, at the minimum of the potential well of the hosting halo and not in a local minimum generated by merging substructures. With this approach, we built a merger tree of the central SMBHs rather than a merger tree of the clusters. We might expect that the two trees differ especially at early epochs (similarly to the small differences between the BCG and the cluster merger trees pointed out in \citealt{2018CINTHIA}). However, we verified that for 80\% of our systems the main progenitor of the SMBH is at the center of the main progenitor of the cluster up to $z=2$ and for half of these the two trees coincide till the time when the SMBH is seeded. \\

\noindent{\it Cluster masses}. Once the center is defined as above, we considered the total gravitational mass of the cluster within an aperture radius $R \leq R_{500}$ computed by summing over all the species of particles: dark matter, cold and hot gas, stars, and black holes. At any redshift we considered only clusters with $M_{\rm cluster}=M_{500} \geq 1.4\times 10^{13} {\rm M}_{\odot}$ or, equivalently, with at least 20 thousands particles within $R_{500}$. The properties of the mass-selected sample are summarized in top part of Table \ref{tab:sample}.\\

\noindent{\it BCG stellar mass}. We defined the mass of the  BCG, $M_{\rm BCG}$, as the stellar mass enclosed in a sphere of radius $0.1 \times R_{500}$ around the cluster center. \\

\noindent \textit{SMBH mass}. Given the identification number of a SMBH particle, the mass of the SMBH, $M_{\rm BH}$, at every redshift is quite easily retrieved from the simulation as it is the mass associated to that particle. The total mass of SMBH particles grows in time because of two separate phenomena: through the accretion of the diffuse gas or via BH-BH mergers. In our simulations, these are the only possible channels for the SMBH to increase its mass. The accretion mass ($M_{\rm BH}^{\rm acc}$) is obtained by integrating the accretion rate, information that we saved at each time step. The merged mass ($M_{\rm BH}^{\rm mer}$) is simply calculated as a difference between the total mass and the accretion mass. As discussed later in the paper, the contribution to the SMBH mass by mergers is negligible at $z \ge 1.5$. Therefore the analysis of this component is restricted to lower redshifts. \\

\noindent{\it Temperature}. In order to compare our results to those from  $\xmm$ observations we considered the spectroscopic-like temperature (\citealt{2004Mazzotta}):
\begin{equation}
	T_{500} = \sum_i \Bigl( \rho_i m_i T_i^{0.25} \Bigr) \Bigl{/} \sum_i \Bigl( \rho_i m_i T_i^{-0.75} \Bigr),
\label{eq2}
\end{equation}
where $\rho_i$, $m_i$, and $T_i$ are the density, mass, and temperature of the $i^{\rm th}$ gas particle within $R_{500}$ emitting in the X-ray band, that is with $T_i > 0.3$ keV and a cold fraction
lower than 10 per cent. We note, in this respect, that according to the effective model by \cite{SPRINGEL2003} adopted in our simulations, the gas particles can be multiphase, carrying information on both hot and cold gas. The cold phase provides a reservoir for stellar formation. In order to have a reliable estimation of the temperature inside $R_{500}$ we imposed two conditions: a minimum of $10^4$ hot gas particles and a maximum fraction of 5 per cent of gas particles discarded because too cold. All clusters satisfying these requirements have also $M_{500}> 1.4\times 10^{13} {\rm M}_{\odot}$, thus whenever we consider measurements of temperature we refer to a subsample of the mass selected-sample. The properties of the temperature-selected subsample are summarized in the bottom part of Table \ref{tab:sample} and the analysis of this subsample is restricted to $z\leq 1$ because at the highest redshift bins, $z=1.5$ and $z=2$, we do not have enough statistics to apply a meaningful analysis. \\

\begin{table}
\caption{Number of clusters, range of mass or temperature covered and their mean values for the mass sample (in the first half) and temperature subsample (in the second half). }
\begin{center}
\begin{tabular}{cccc}
\hline
& & $M$ sample \\
 $z$ & $N$ & $M_{500}$ [$10^{14}$M$_{\odot}$] & $<M>$ [$10^{14}$M$_{\odot}$] \\
 0.0 & 135 & 0.14-25.83 & 2.94  \\
 0.5 & 114 & 0.14-14.11 & 1.65  \\
 1.0 & 85 & 0.14-5.15 & 0.99  \\
 1.5 & 59 & 0.14-2.48 & 0.69  \\
 2.0 & 37 & 0.15-1.59 & 0.52  \\
\hline
& & {$T$ subsample}\\
$z$ & $N$ & $T_{500}$ [keV] & $<T>$ [keV] \\
 0.0 & 93 & 0.80-10.81 & 3.27  \\
 0.5 & 62 & 0.80-8.89 & 2.97  \\
 1.0 & 35 & 0.95-5.84 & 3.01  \\
\hline
\end{tabular}
\end{center}
\label{tab:sample}
\end{table}

\noindent{\it Best-fitting procedure.} For all the considered relations, we looked for a best-fit line in the logarithmic plane:
\begin{equation}
\log(Y/Y_0)=a+b \times \log(X/X_0),
\end{equation}
where $\log$ always indicates the decimal logarithm. The temperature, cluster mass, BCG stellar mass, and SMBH mass are always normalized by the same factors, expressed above as $X_0$ or $Y_0$ and respectively equal to 2 keV, $10^{14} {\rm M}_{\odot}$, $10^{11} {\rm M}_{\odot}$, and $10^{9} {\rm M}_{\odot}$.

To find the best-fit curve, we employed an IDL routine that is resistant with respect to outliers: ROBUST\_LINEFIT\footnote{https://idlastro.gsfc.nasa.gov/ftp/pro/robust/robust\_linefit.pro}. For the simulated data, we always considered the BISECT option, recommended when the errors on X and Y are comparable so there is no true independent variable. This is particularly appropriate in the case of numerical simulations where no errors are linked to measurements. 
To estimate the error associated with the parameters of the best-fitting relation, we generated ten thousand bootstrap samples by randomly replacing the data. From the resulting distributions we derived the mean values and the standard deviations to be associated, respectively, with the parameters and their errors. All relevant best-fitting coefficients of the linear regressions are reported in Table~\ref{tab:all} and will be discussed in the next two sections.

\section{Results}\label{results}

\subsection{Comparison with observational data}\label{Comparison with observational data}
In this section we compare the numerical results to the observations presented in \cite{2018BOGDAN}, where the correlation between the mass of the SMBHs in BCG and the global temperature of clusters and groups of galaxies is presented. We also complement this analysis with the more recent data of \cite{2019GASPARI}\footnote{This sample includes not only massive galaxies in groups and clusters, but also isolated and spiral galaxies.}. \cite{2018BOGDAN} derive the temperature from \textit{XMM-Newton} observations of the hot gas; \cite{2019GASPARI} use published data of Chandra and wide-field \textit{ROSAT/XMM Newton}. The extraction region for both is on average $\sim0.2\,R_{500}$ (core included; such temperature is a good proxy for $T_{500}$) .

\begin{figure}
	\includegraphics[width=\columnwidth]{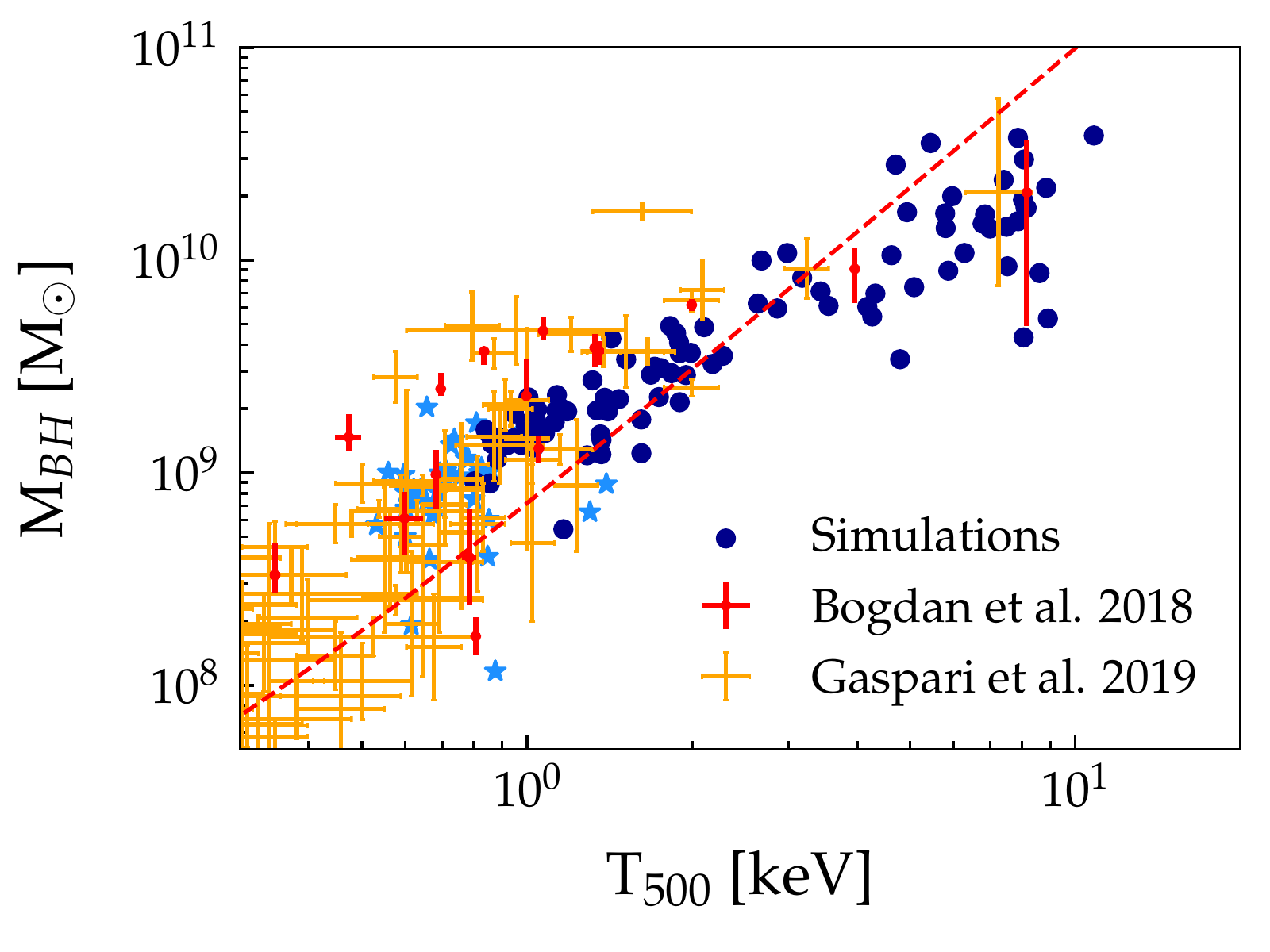}
    \caption{Correlation between SMBH mass, $M_{\rm BH}$, and the clusters temperature, $T_{500}$. Red and orange crosses refer to observational data from \protect\cite{2018BOGDAN} and \protect\cite{2019GASPARI} respectively. Dark-blue dots represent simulated clusters in the temperature subsample while cyan stars show the remaining objects of the mass sample. Dashed red line is the prediction of the toy model by \protect\cite{2005CHURAZOV}.}
    \label{fig:T_mbh}
\end{figure}

In Fig.~\ref{fig:T_mbh} we show the correlation between $M_{\rm BH}$ and $T_{500}$ for our simulations and for their observational data set. In the figure we also show as a dashed red line the results of the simple toy model by \cite{2005CHURAZOV} \footnote{Equation $7$ of \cite{2005CHURAZOV} with $\delta_E=10^{-4}$, $t_9=1$, $\Lambda(T)$ as defined in \cite{2001TOZZI} with $Z=0.3\ Z_{\odot}$, and $T=3\times 10^6\ \sigma_{200}\ K$}, which even under simplified assumptions reproduces the observed correlation. We find a good agreement between observations and simulations. Nevertheless, a more quantitative comparison between the two samples is difficult for an under-representation of clusters with $T_{500}>2$ keV in the observational sample that, however, has a good number of systems below 1 keV. 

\begin{table}
\caption{Parameters of all best-fitting lines derived with the procedure described at the end of Sect.~3. For each (X,Y) pair of data, we fit the formula: $\log(Y/Y_0)=a+b\times \log(X/X_0)$ where the normalizations, $X_0$ or $Y_0$, are equal to 2 keV, $10^{14} M_{\odot}$,$10^{11} M_{\odot}$ and $10^9 M_{\odot}$, respectively, for $T, M_{500},M_{\rm BCG}$, and all SMBH masses: $M_{\rm BH},M_{\rm BH}^{\rm mer}$, and $M_{\rm BH}^{\rm acc}$. The parameters $a, b$, and $\sigma$ and their errors are the mean and standard deviation values of the distributions obtained by applying the procedure to ten thousand bootstrapping samples. The asterisks indicate that the analysis is performed to the temperature subsample. }
\begin{tabular}{lcccc}
\hline
 $(X,Y)$ & $a$  & $b$   & $\sigma_{Y|X}$  \\
\hline

$z=0$\\
$M_{500},T^*$                    & $0.10\pm0.01$ & $1.71\pm0.03$ & $0.07\pm0.01$  \\
$T,M_{\rm BH}^*$                 & $0.52  \pm 0.02$ & $1.28 \pm 0.06$ & $0.16 \pm 0.02$  \\
$M_{500},M_{\rm BCG}$            & $0.75  \pm 0.01$ & $0.66 \pm 0.01$ & $0.10 \pm 0.01$  \\ 
$M_{\rm BCG},M_{\rm BH}$         & $-0.42 \pm 0.03$ & $1.16 \pm 0.04$ & $0.14 \pm 0.01$ \\
$M_{500},M_{\rm BH}$           & $0.45  \pm 0.02$ & $0.76 \pm 0.03$ & $0.18 \pm 0.02$ \\
$M_{500},M_{\rm BH}^{\rm mer}$ & $0.20  \pm 0.02$ & $0.73 \pm 0.04$ & $0.23 \pm 0.02$  \\
$M_{500},M_{\rm BH}^{\rm acc}$ & $0.03  \pm 0.02$ & $0.83 \pm 0.04$ & $0.25 \pm 0.02$  \\
 \\
$z=0.5$\\ 
$M_{500},T^*$                    & $0.05 \pm 0.01$ & $1.73 \pm 0.04$ & $0.08 \pm 0.01$  \\
$T,M_{\rm BH}^*$                 & $0.48 \pm 0.03$ & $1.49 \pm 0.11$ & $0.24 \pm 0.03$  \\
$M_{500},M_{\rm BCG}$            & $0.68 \pm 0.01$ & $0.69 \pm 0.02$ & $0.13 \pm 0.01$  \\
$M_{\rm BCG},M_{\rm BH}$         & $-0.42\pm 0.03$ & $1.25 \pm 0.05$ & $0.16 \pm 0.02$ \\ 
$M_{500},M_{\rm BH}$           & $0.43 \pm 0.02$ & $0.86 \pm 0.04$ & $0.23 \pm 0.02$  \\
$M_{500},M_{\rm BH}^{\rm mer}$ & $0.06 \pm 0.03$ & $0.92 \pm 0.06$ & $0.34 \pm 0.04$  \\
$M_{500},M_{\rm BH}^{\rm acc}$ & $0.15 \pm 0.03$ & $0.90 \pm 0.06$ & $0.27 \pm 0.03$ \\
 \\
 $z=1$\\ 
$M_{500},T^*$                    &  $-0.07 \pm 0.01$ & $1.72\pm 0.05$ & $0.06 \pm 0.01$  \\
$T,M_{\rm BH}^*$                 &  $0.46 \pm 0.04$ & $1.70 \pm 0.22$ & $0.22 \pm 0.03$ \\
$M_{500},M_{\rm BCG}$            &  $0.67 \pm 0.01$ & $0.73 \pm 0.03$ & $0.11 \pm 0.01$ \\
$M_{\rm BCG},M_{\rm BH}$         &  $-0.42\pm 0.04$ &  $1.45\pm 0.07$ & $0.19\pm 0.02$ \\
$M_{500},M_{\rm BH}$           &  $0.56 \pm 0.04$ & $1.06 \pm 0.08$ & $0.27 \pm 0.03$ \\
$M_{500},M_{\rm BH}^{\rm mer}$ &  $0.12 \pm 0.04$ & $1.21 \pm 0.17$ & $0.40 \pm 0.09$  \\
$M_{500},M_{\rm BH}^{\rm acc}$ &  $0.35 \pm 0.05$ & $1.07 \pm 0.09$ & $0.30 \pm 0.03$   \\
\\
$z=1.5$\\  
$M_{500},M_{\rm BCG}$            & $0.69 \pm  0.02$ & $0.77 \pm 0.05$& $0.12 \pm 0.02$  \\
$M_{\rm BCG},M_{\rm BH}$          & $-0.38 \pm 0.05$ & $1.58 \pm 0.10$ & $0.22 \pm 0.02$ \\ 
$M_{500},M_{\rm BH}$           & $ 0.71 \pm 0.05$ & $1.25 \pm 0.11$ & $0.30\pm 0.04$  \\
$M_{500},M_{\rm BH}^{\rm acc}$ &  $0.55  \pm 0.06$ & $1.27 \pm  0.11$ &  $0.33 \pm  0.05$\\
\\
$z=2$ \\
$M_{500},M_{\rm BCG}$             & $0.67 \pm 0.02$ & $0.75\pm 0.06$ & $0.11 \pm 0.03$    \\
$M_{\rm BCG},M_{\rm BH}$          & $-0.33 \pm 0.08$ & $1.76 \pm 0.20$ & $0.26 \pm 0.05$ \\ 
$M_{500},M_{\rm BH}$           & $0.85 \pm 0.08$ & $1.35\pm 0.15$ & $0.34 \pm 0.04$   \\
$M_{500},M_{\rm BH}^{\rm acc}$ & $0.72 \pm 0.10$ & $1.37\pm 0.16$ & $0.36 \pm 0.04$ \\
 \\ 
\hline
\label{tab:all}
\end{tabular}
\end{table}
In order to better populate the colder and less massive tail of the simulated cluster distribution we compare the correlation between $M_{\rm BH}$ and $M_{500}$ by using the mass sample rather than the less-numerous temperature sub-sample (Table~1). In \cite{2018BOGDAN} the total mass was not measured directly from their data but was derived from the temperature via the scaling relation by \cite{2015LOVISARI}:
\begin{equation}
	M_{500} = 1.11 \times 10^{14} (kT/\text{2 keV})^{1.65} \text{M}_{\odot} .
    \label{eq:lovi}
\end{equation}

\begin{figure}
	\includegraphics[width=\columnwidth]{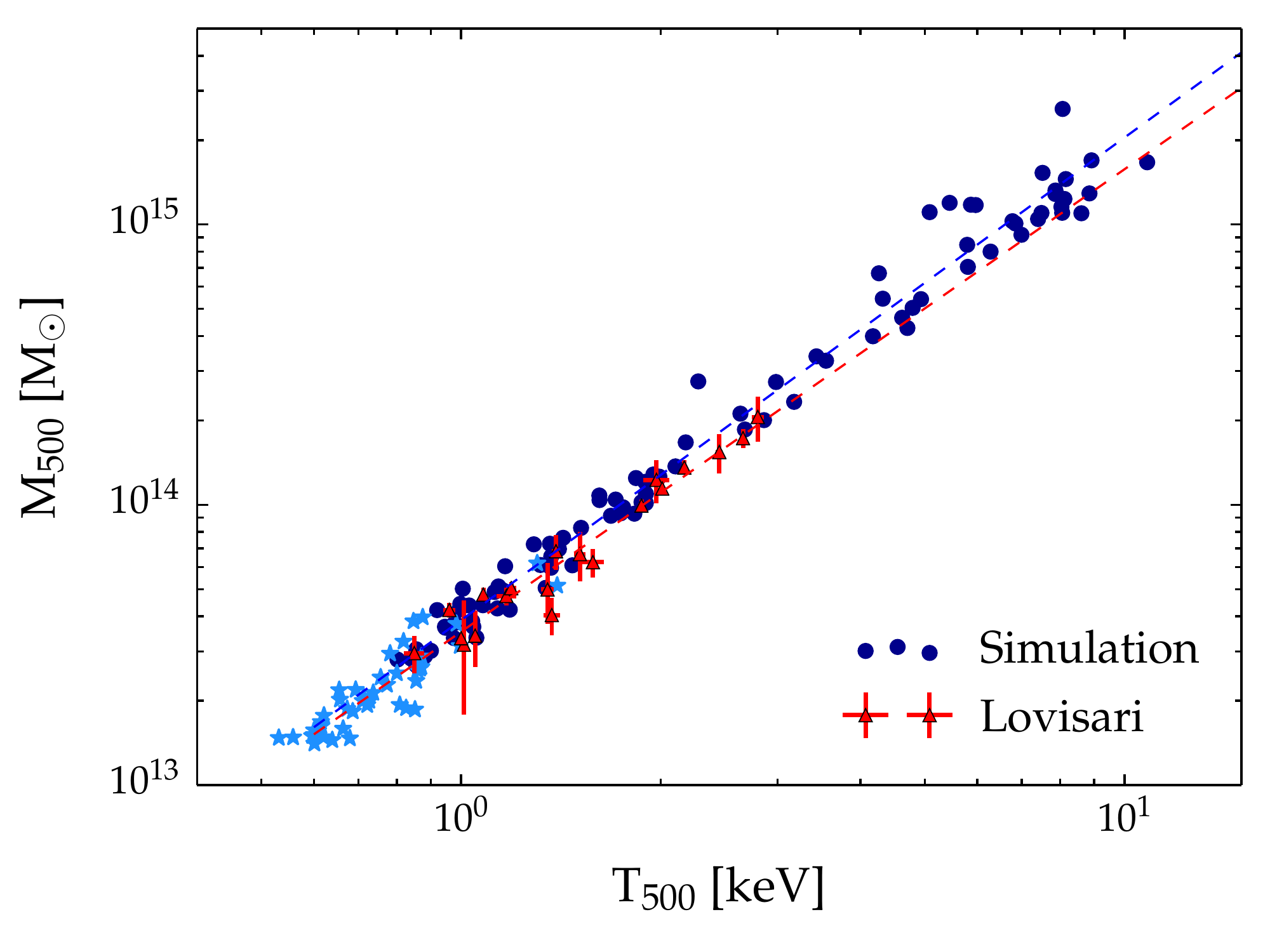}
    \caption{Correlation between cluster mass, M$_{500}$, and cluster temperature, T$_{500}$. Symbols as in Fig.\ref{fig:T_mbh}, where the observational data are taken from \protect\cite{2015LOVISARI}. Dashed lines are the best-fitting lines for both simulated and observed data.} 
    \label{fig:M-T_relation}
\end{figure}

For this reason, before analyzing the $M_{\rm BH}$-$M_{500}$ relation, it is helpful to compare the observed and simulated $T_{500}$-$M_{500}$ relations. The observed and simulated data sets are shown in Fig. \ref{fig:M-T_relation} together with the best-fitting linear relations. In case of the observed sample we verified that our fitting procedure, without the BISECT option,
returned the same parameters of Eq.~\ref{eq:lovi}. In particular we confirm the value of the slope reported in \cite{2015LOVISARI}: $b=1.65 \pm 0.07$. 

By looking at Fig.~\ref{fig:M-T_relation}, we see a good agreement between simulated and observed clusters in the temperature range covered by \cite{2015LOVISARI}. However, the extrapolation of their best-fit line suggests a possible difference in the hottest-clusters regime.  The two slopes agree within 1$\sigma$, but the observed clusters have on average slightly higher temperature with respect to simulated clusters at fixed mass. For example, the temperature of observed clusters is $9$  per cent higher at $M_{500}=10^{14} {\rm M}_{\odot}$. This feature is not new and has been already noted in \cite{2018TRUONG}, where a similar set of numerical simulations is employed, and, more interestingly, in other numerical analysis, such as \cite{FABLE}. In particular, in their work the authors show that numerical results are in agreement with observations if are considered only cluster masses derived via weak lensing. This suggests that  the observed X-ray hydrostatic masses are biased low.

\begin{figure}
	\includegraphics[width=\columnwidth]{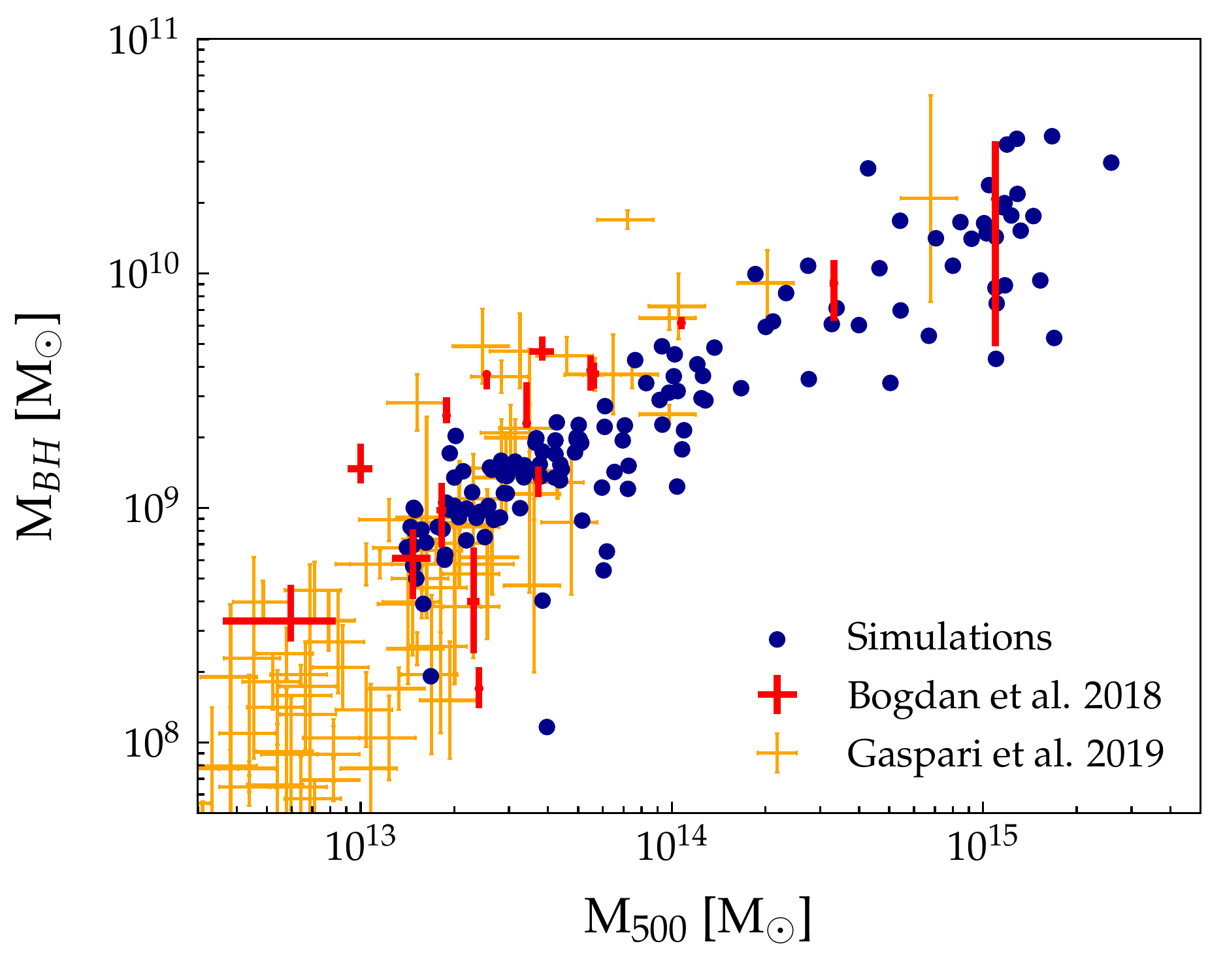}
    \caption{Correlation between SMBH mass, M$_{\rm BH}$, and cluster mass, $M_{500}$.  Symbols as in Fig.\ref{fig:T_mbh}.} 
    \label{fig:m500_mbh}
\end{figure}

Finally, in Fig.~\ref{fig:m500_mbh} we compare the correlation between $M_{\rm BH}$ and the $M_{500}$ as measured in our simulations and as derived in \cite{2018BOGDAN} and \cite{2019GASPARI}. The results of the comparison are expected from the previous two figures: the simulated data points are in line with observations, especially at high ($M_{500}> 3 \times 10^{14} {\rm M}_{\odot}$) and low masses ($M_{500}<3\times 10^{13}{\rm M}_{\odot}$). In the intermediate mass range, the few observed data points tend to have slightly higher SMBH masses than the simulated objects. This apparent mis-match is presumably a consequence of the poor statistics of $10^{14} {\rm M}_{\odot}$ objects in the observational sample. More unlikely, this feature could suggest a broken power law to describe the $M_{\rm BH}-M_{500}$ relation, but such a drastic change in the slopes is difficult to justify. 

\subsection{The theoretical $M_{\rm BH}$-$M_{500}$ relation}

\begin{figure}
	\includegraphics[width=\columnwidth]{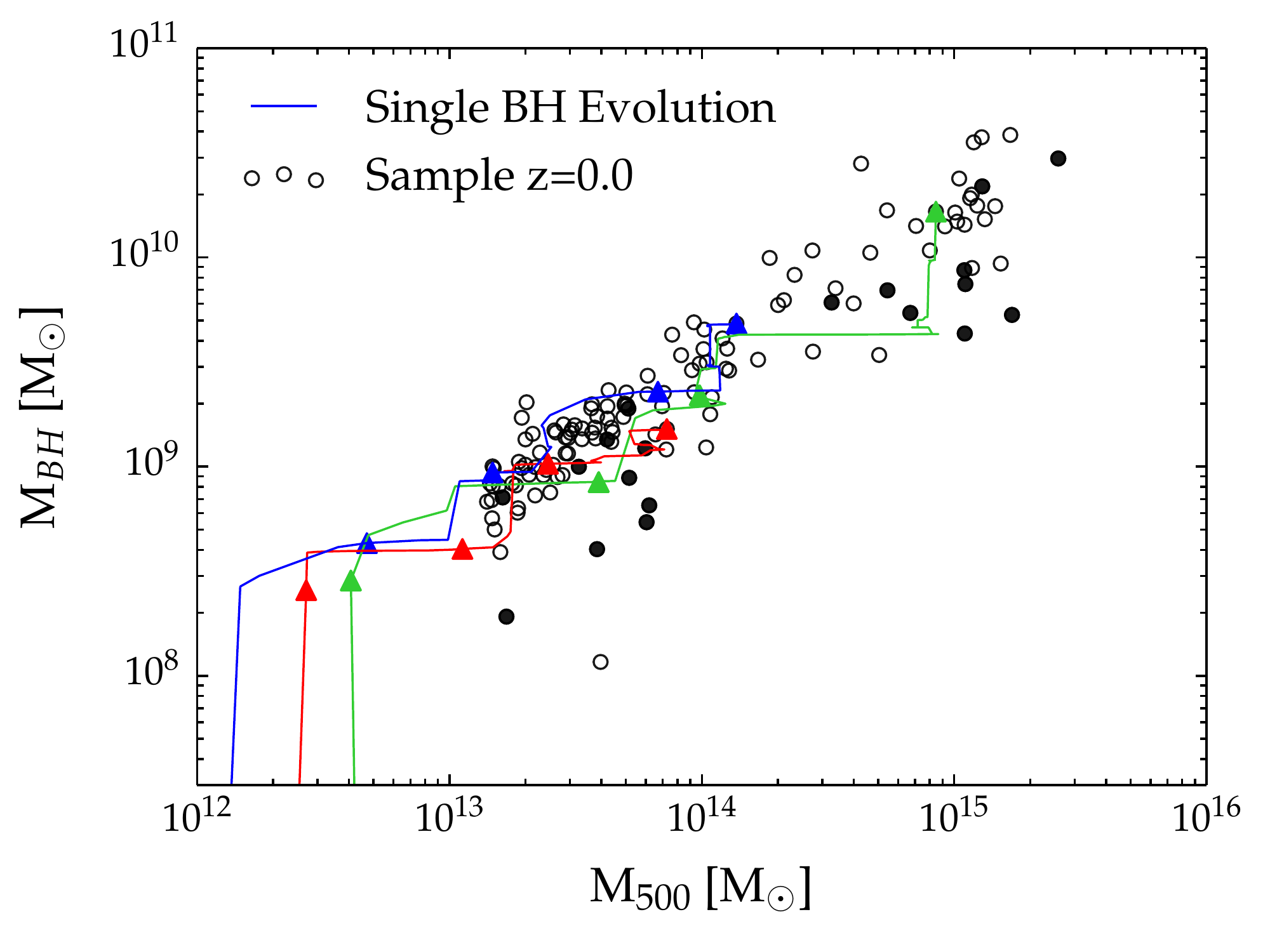}
    \caption{Evolutionary tracks of three different systems on the $M_{\rm BH}$-$M_{500}$ plane. Triangles over each line indicate the position of the systems on the plane at $z=3$, $z=2$, $z=1$ and $z=0$. Black circles represent our numerical sample at $z=0$; the filled ones are systems for which $M_{500}$ is increased by more than 40 percent in the last Gyr.}
    \label{fig:mbh_single_evolution}
\end{figure}

Since the simulated sample is in overall good agreement with the observed correlation between the mass of the central SMBHs and the mass of the clusters, we investigate here how single simulated systems evolve throughout time to form, by $z=0$, the $M_{\rm BH}-M_{500}$ relation shown in Fig.\ref{fig:mbh_single_evolution}.
For this goal, we over-plotted three evolutionary tracks of representative SMBHs. These objects are chosen accordingly to their mass at $z=0$; specifically, they refer to a small, a medium-mass, and a massive SMBH. To have some temporal reference we also indicated four specific times along each line: $z=3$, $z=2$, $z=1$, and $z=0$.

Despite the different final masses, the evolutionary tracks of the three systems have strong similarities which are common also in all the other objects analyzed (not shown for sake of clarity). Three phases are clearly distinguishable. 
At the highest redshifts, the mass of the SMBHs grows rapidly at almost constant $M_{500}$. 
This track begins instantaneously as the SMBHs are seeded in a gas rich region. The SMBHs immediately gain mass at the Eddington limit by the accretion from the abundant surrounding gas, which is mostly cold and thus efficient at increasing the SMBH mass. This phase typically lasts half Gyr and can lead to the formation of SMBHs with a mass already of the order of $M_{\rm BH} \approx 10^8-10^9 {\rm M}_{\odot}$,  in line with other hydrodynamical cosmological simulations (e.g., \citealt{2018MNRAS.481.3118M}). The fast SMBH growth ends when the $M_{\rm BH}$ is high enough to cause an intense feedback that leads to the ejection of part of the gas outside the shallow potential wells of the hosting galaxies. By then, all SMBHs in our sample are close to the $M_{\rm BH}-M_{500}$ relation. This happens before $z=2$ and in some cases even at $z>4$.  

After this initial phase, the cluster and its central SMBH co-evolve, but not with a linear evolutionary track: the increase of the SMBH and cluster masses is not simultaneous. The shape of the tracks, instead, can be described as a stairway: the systems evolve in this plane either at almost constant $M_{\rm BH}$ or at almost constant $M_{500}$. The former situation occurs during cluster mergers. It starts when merging structures reach and cross $R_{500}$ leading to a quick increment of the total cluster mass and finishes when the secondary objects are fully incorporated. These horizontal shifts in the $M_{\rm BH}$-$M_{500}$ plane typically last 1 Gyr or less and only in the rare cases when multiple mergers are subsequently taking place they can last up to 2 Gyr. In the following period, spanning from 1 to 3 Gyr, the substructures move towards the center of the cluster and neither the SMBH mass nor the cluster mass change. Eventually, the merging objects reach the core and either feed the central SMBH with gas or induce a BH-BH merger or both. The event is captured by the vertical movement in the plot. 

All these time-frames are clearly indicative as they depend on several parameters that characterize the merger events such as the mass ratio and the impact parameter. Nonetheless, it is always the case that the mass of the SMBH and of the cluster are for the largest majority of time at the connection between the horizontal and vertical steps rather than along their tracks. This behavior indicates that the scatter of the relation might differ when the sample is selected according to the dynamical status of the SMBH hosts. We expect that the points related to relaxed BCG in relaxed clusters will always be above the points linked to systems where either the BCG or the cluster are experiencing, or just experienced, a merger event. Indeed, we expect that relaxed and perturbed systems are respectively located in our plot on the top and the bottom of the vertical segments. In favor of this picture, it is noticeable that the 20 clusters whose $M_{500}$ grows by more than 40 per cent in the last Gyr (shown as black points in Fig. \ref{fig:mbh_single_evolution}) have SMBHs that on median are 50 per cent smaller than those expected  to follow the M$_{500}$-M$_{\rm BH}$ relation.

\subsection{Evolution of the M$_{\rm BH}$-M$_{500}$ relation}
\begin{figure}
 \includegraphics[width=\columnwidth]{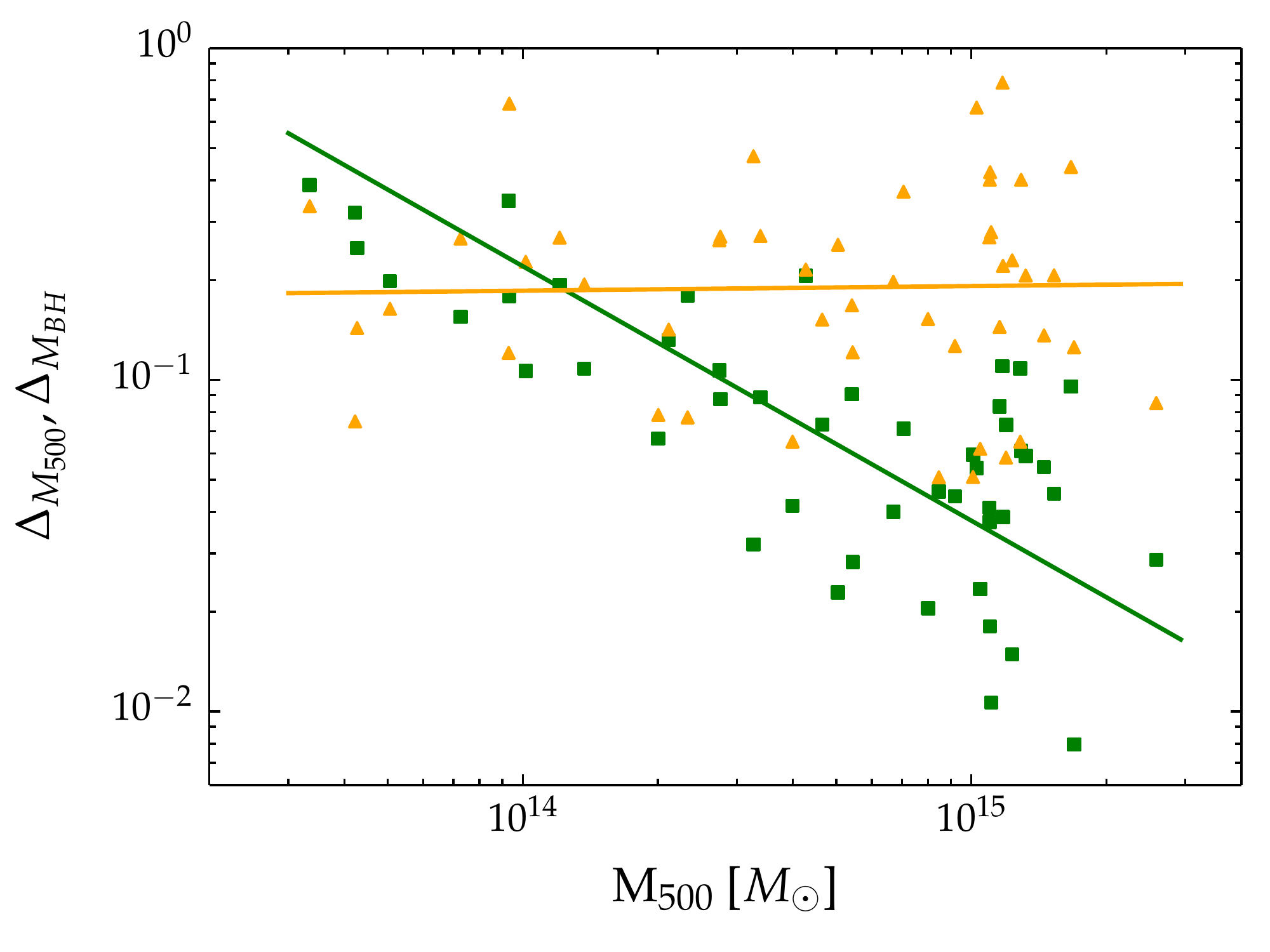}
\caption{Ratios between cluster (SMBH) masses at $z=2$ and cluster (SMBH) masses at $z=0$ as a function of the cluster masses at $z=0$. Clusters are shown as green squares and SMBHs as yellow triangles. The sample used is the mass-selected sample identified at $z=2$. }
    \label{fig:deltam500}
\end{figure}

After the inspection on the trajectory of individual objects we study here the evolution of the entire M$_{500}$-M$_{\rm BH}$ relation. We start with the evaluation of the ratio between the mass gained by clusters and by central SMBHs between $z=2$ and $z=0$: $\Delta_M=M_{z=2}/M_{z=0}$, where $M$ refers to either the total mass, $M_{500}$, or the SMBH mass, $M_{\rm BH}$. If these two ratios are constant, the slope of the relation will not change. The resulting ratios are shown in Fig.~\ref{fig:deltam500} as a function of the cluster total mass reached at $z=0$ for the mass sample identified at $z=2$.
From the plot, we can clearly infer that the variation in total mass between the two epochs is strongly mass dependent. The absolute value of the slope of the best-fitting $\Delta_{M_{500}}-M_{500}$ relation is, indeed, greater than 0.75. Clusters with a final mass lower than $10^{14}{\rm M}_{\odot}$ increase their total masses by a factor between 3 and 6. Instead, massive clusters with final $M_{500} > 10^{15} {\rm M}_{\odot}$ increase their mass on average by a factor of about $30$ with individual objects that can grow by more than two orders of magnitude. This feature is completely in line with the expectations of hierarchical clustering.  

The high mass regime is particularly characterized by a large spread that is representative of what we might expect from a volume-limited sample because the most massive objects, $M_{500} > 10^{15}$ M$_{\odot}$ correspond to the most massive systems of the parent volume-limited cosmological box (see Sect.~2). Vice versa, the scatter for the smallest systems is likely under-estimated. Indeed,  the linear trend is expected to flatten for the lowest masses to a constant growth rate value.
On the other side, Fig.~\ref{fig:deltam500} also shows that the variation on the SMBHs mass is independent of the cluster mass and that SMBHs grow on average by a factor of about 5-6. As a consequence we expect a marked evolution of the slope of the $M_{\rm BH}$-$M_{500}$ relation between $z=2$ and $z=0$. 

\begin{figure}
	\includegraphics[width=\columnwidth]{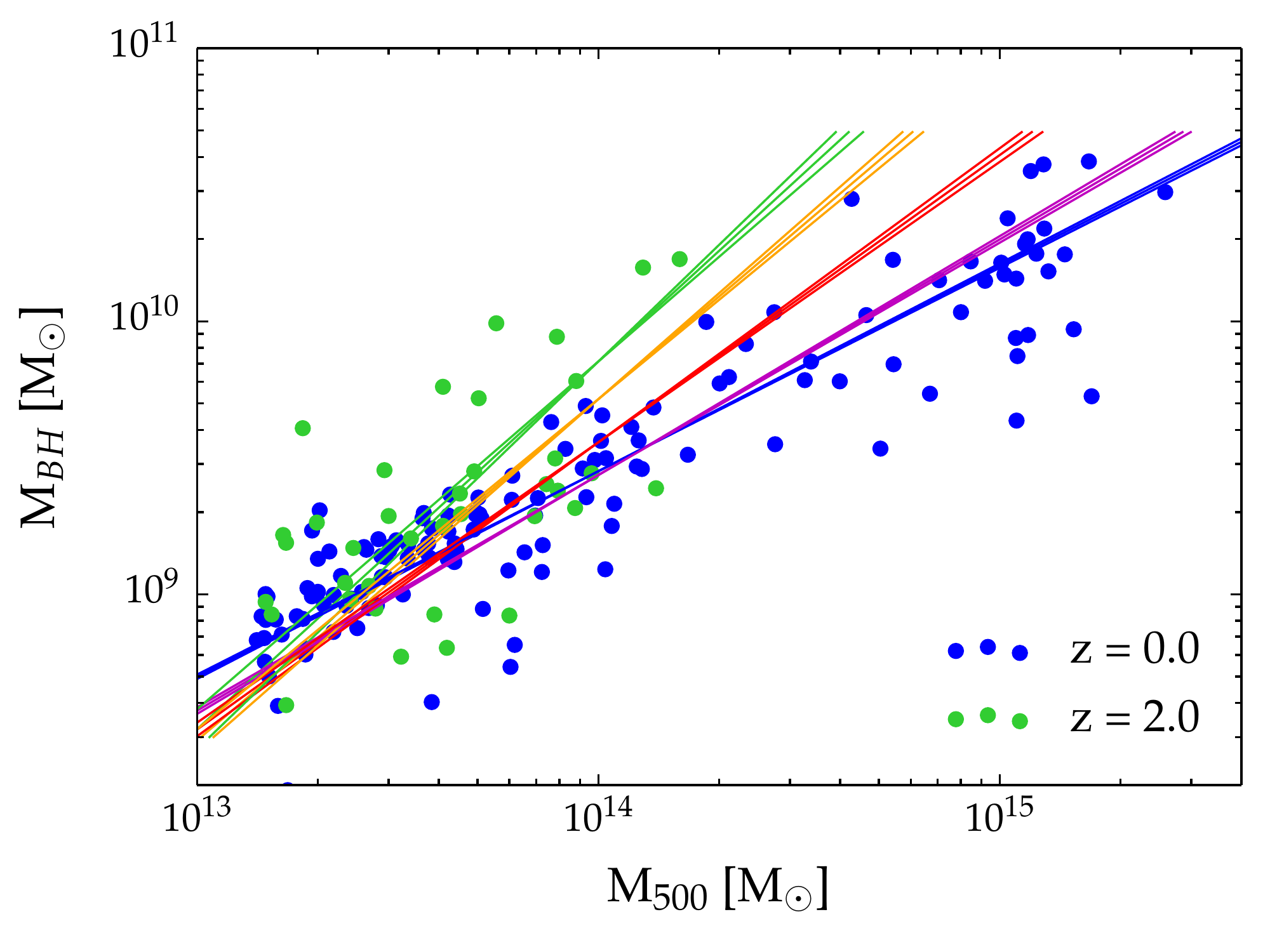}
    \caption{Correlation between M$_{\rm BH}$ and M$_{500}$ at different redshifts. At every redshift we show the best-fitting relation in the mass range of the respective samples. Namely, we show in green, orange, red, magenta, and blue the mass sample related to $z=2, 1.5, 1, 0.5$ and $z=0$, respectively.}
    \label{fig:m500_mbh_evolution_whole_sample}
\end{figure}

This is confirmed in Fig. \ref{fig:m500_mbh_evolution_whole_sample} where we plot the best-fitting lines for our mass samples at all redshifts considered. 
The relations are steeper at higher redshifts: the value at $z=2$, $b=1.348$, is almost twice the value found at $z=0$, $b=0.753$. From Fig. \ref{fig:deltam500} and Fig. \ref{fig:m500_mbh_evolution_whole_sample} we conclude that the change in the slope is mainly driven by the different evolution rate of the most massive clusters with respect to the smallest objects, trend that is in line with the expectations from $\Lambda$CDM cosmology. 

\subsection{Evolution of SMBH mass}

\begin{figure}
	\includegraphics[width=\columnwidth]{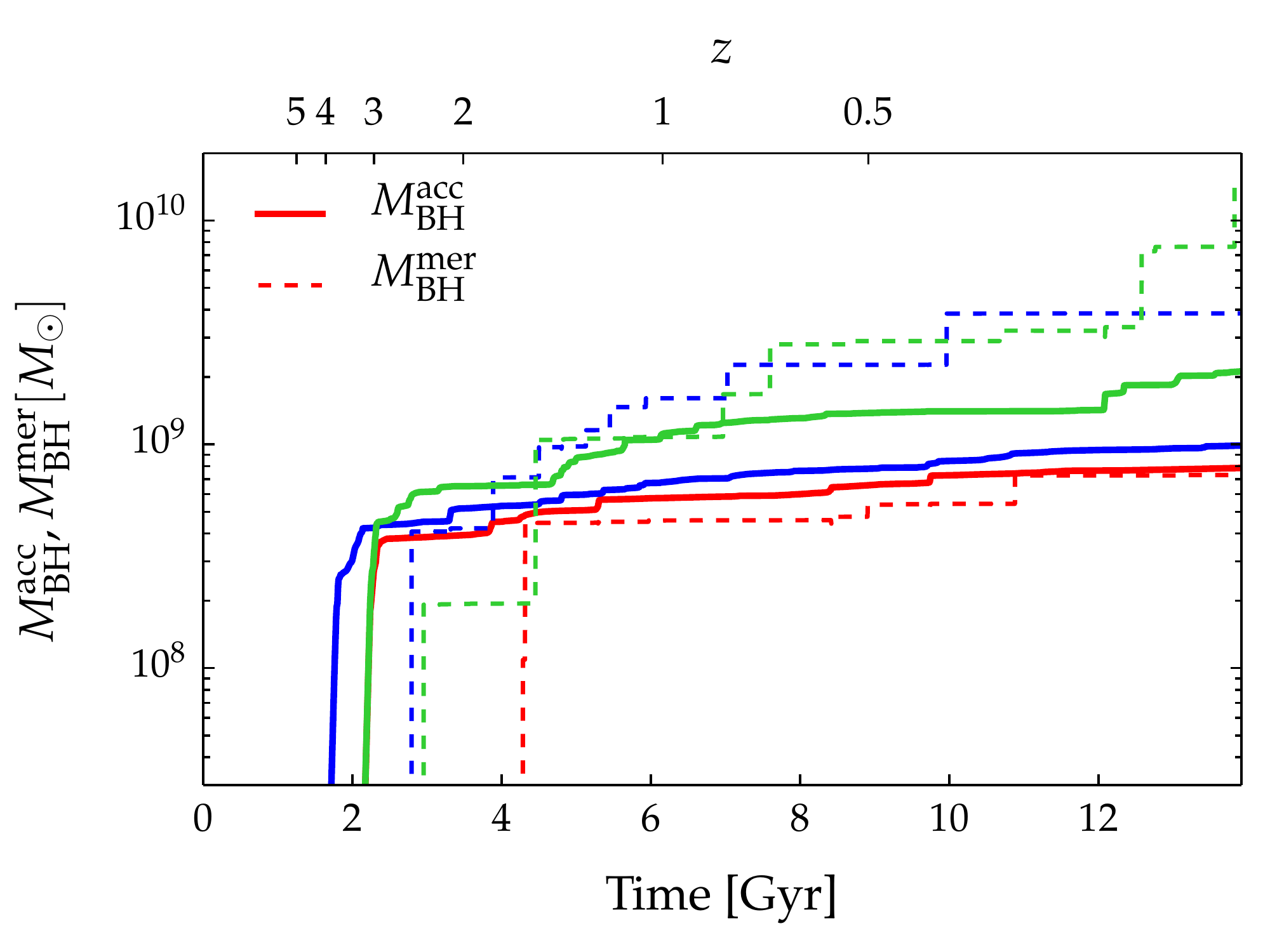}
    \caption{Evolution of the mass of black holes showed in Fig. \ref{fig:mbh_single_evolution}. The solid lines represent the mass of the black holes due to gas accretion. Dashed lines represent mass gained via BH-BH mergers. }
    \label{fig:accretion_merger_evolution}
\end{figure}
\begin{figure}
	\includegraphics[width=\columnwidth]{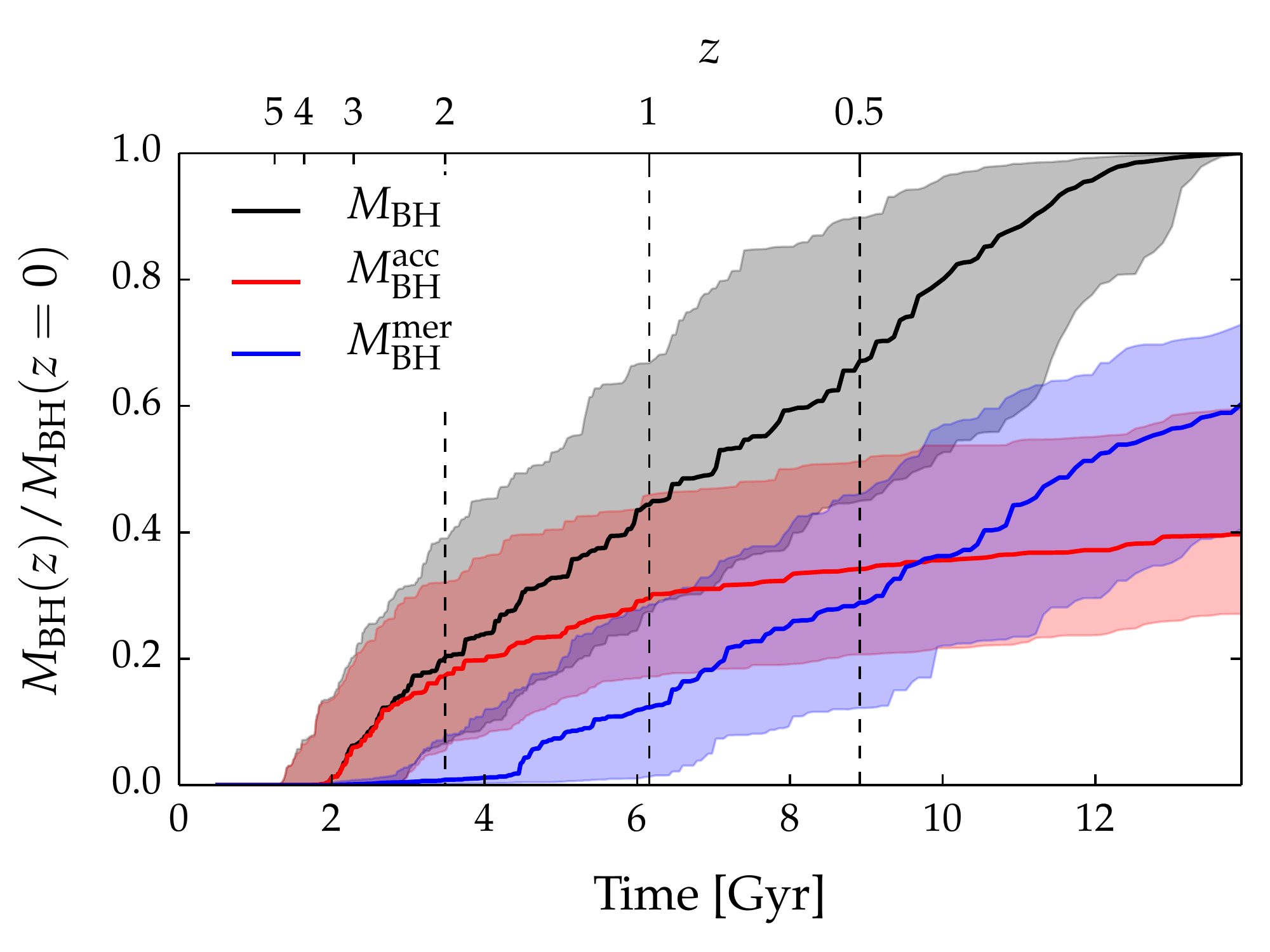}
    \caption{Evolution of SMBH mass divided into two growth channels (gas accretion in red and BH-BH mergers in blue) considering the complete sample. In black we also plotted the total SMBH mass. Solid lines represent median values of our sample and shadowed regions represent 16 and 84 percentiles.}
    \label{fig:histo_accretion_merger}
\end{figure}

To better understand how the black hole mass evolves with time we separately study the two mechanisms that contribute to the growth of the SMBH mass: the accretion by the diffuse gas, $M_{\rm BH}^{\rm acc}$, and the merger with other SMBHs, $M_{\rm BH}^{\rm mer}$. First, we  analyzed how the mass of single SMBH evolves with time via the two separate channels. As an example in Fig. \ref{fig:accretion_merger_evolution} we plot the evolution of the three SMBHs shown in Fig. \ref{fig:mbh_single_evolution}. As already noted before, the evolution is characterized by an initial phase of intense gas accretion that for these three systems is approximately between $z=5$ and $z=3$. After that, SMBHs still grow by gas accretion but at a much smaller rate. On the contrary the increase of the SMBH mass due to BH-BH mergers becomes more important and it is the main channel of the SMBH mass growth at lower redshifts, that is $z\leq 1$ for the two most massive SMBHs and $z\leq 0.5$ for the smallest one.

In Fig. \ref{fig:histo_accretion_merger} we show the evolution of our complete sample. We plot the median behaviors with a solid line and the 68 per cent of the total sample distributions (from the 16$^{\rm th}$ to the 84$^{\rm th}$ percentiles) with the shaded regions. The total SMBH mass and the masses gained from the two channels are normalized with respect to the total SMBH mass at $z=0$. Finally, the dashed vertical lines help to identify three significant times: $z=0.5$, 1, and 2. 

From the SMBH seeding up to $z\approx 2$ the total mass of the SMBH grows almost entirely by gas accretion. Half of the final mass gained through gas accretion is, indeed, accumulated before $z=2$. Then from $z\sim 2$ to $z\sim 1$ the mass growth due to BH-BH merger becomes more relevant increasing at a rate comparable to the growth rate of $M_{\rm BH}^{\rm acc}$. By $z=1$ $M_{\rm BH}^{\rm mer}$ makes up on average 25 per cent of the total mass at that redshift. At lower redshift, $z<0.5$,  BH-BH mergers provide the main channel for SMBH mass growth, and eventually $M_{\rm BH}^{\rm mer}$ represents the main component of mass gained by $z=0$, in line with some previous results (e.g., \citealt{2013MARTA}; \citealt{2014DUBOIS}; \citealt{2018MNRAS.479.4056W}). Indeed, the mass accumulated by gas accretion from $z=1$ to $z=0$ accounts only for $10$ per cent of the total final mass, while during the same period the SMBH gains $50$ per cent of its final mass via mergers.
\begin{figure}
	\includegraphics[width=\columnwidth]{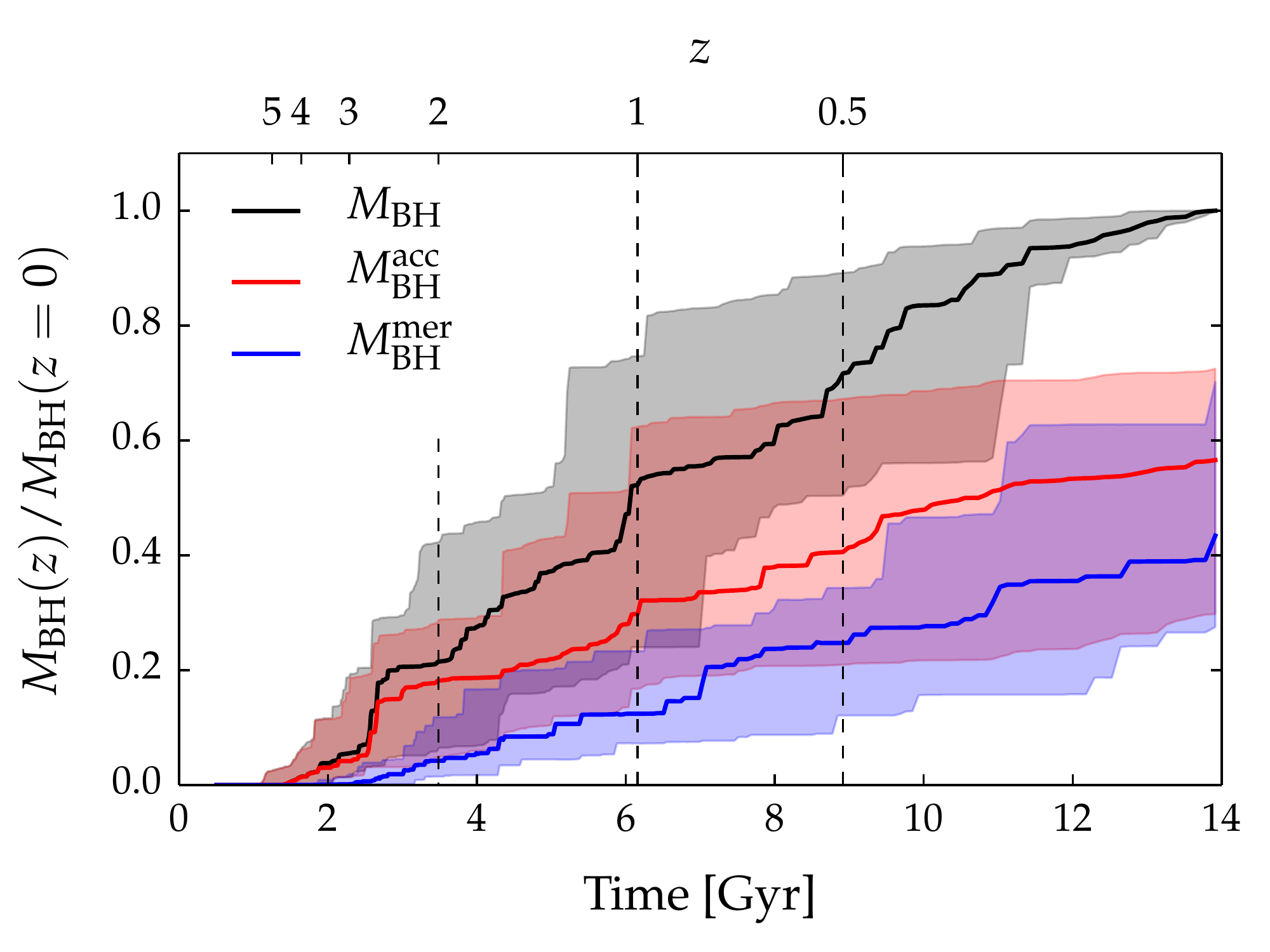}
	\caption{Same as Fig.~9 but considering only clusters with $M_{500}>10^{15} M_{\odot}$ at $z=0$.}
    \label{fig:massive}
\end{figure}

\vspace{0.5cm}

The relative importance of the two channels shown in the figure is, however, characterized by a large scatter. We, therefore, explored whether this is due to the broad mass range investigated and, thus, whether the described behavior depends on the mass of the systems. We divided the sample in three mass bins: at $z=0$ the least massive objects have $M_{500}$ below $10^{14} M_{\odot}$, the most massive above $10^{15} M_{\odot}$, and the intermediate in between these two thresholds. We found that the trends of the relative ratio of the two SMBH growth channels are extremely similar to Fig.~\ref{fig:histo_accretion_merger} for the samples of the smallest and intermediate objects. This result is expected for the first mass bin since it is the most numerous, containing 84 objects, but it was not guaranteed for the intermediate sample with only 31 systems. The most massive sample, however, is on average characterized by a continuous and equivalent growth of both channels after $z=2$ (see Fig.~10). As a result, at $z=0$ $M_{\rm BH}^{\rm acc}$ accounts for $\approx 60 $ per cent of $M_{\rm BH}$. That said, we also notice that the scatter remains very large and the distributions related to the two channels show a large intersecting area. We, therefore, can conclude that the scatter shown in Fig.~\ref{fig:histo_accretion_merger} is not related to the total mass of the systems.

The stronger relative influence by the SMBH accretion with respect to the BH-BH merger in most massive clusters can be due to two different factors: on the one hand AGN feedback is not able to completely balance gas cooling, on the other hand BH-BH mergers could be less frequent. In the following we show some evidence that both phenomena are actually in place.

To demonstrate that the AGN are less efficient in regulating the gas cooling in the cores of massive clusters we computed the total energy released by AGN feedback for $z<1$ and related it to the gas mass within $0.1 \times R_{500}$. We find that the ratio of the two quantities is a strongly decreasing function of $M_{500}$ and that it changes by more than a factor of 10 from the least massive to the most massive systems. This suggests that the heating provided by the AGN feedback is relatively smaller for large objects where, therefore, the gas cooling is less contrasted. The central SMBH has therefore more cold gas at its disposal.

To test the reduced frequency of BH-BH merger, we computed at $z=0$ the number and mass of SMBHs which are inside $0.1 \times R_{500}$ and are not bound to any substructure. We find that 80 per cent of the most massive systems (16 objects over 20) have several SMBHs in that central regions with a total mass greater than 10 percent of the mass of the central SMBH. Analysing the first mass bin ($M_{500} < 10^{14} M_{\odot}$), instead, only 10 per cent of clusters have enough SMBHs able to account, all together, for at least 10 per cent of the mass of the central SMBH. This is mostly due to the fact that more massive clusters host more massive and extended BCGs. When the substructures interact with these well-established BCGs they are more easily disrupted (see Sect.~4.5 for further details) at a larger radii, preventing or delaying mergers between their SMBHs and the central SMBH. 
%
\begin{figure*}
	\includegraphics[width=\columnwidth]{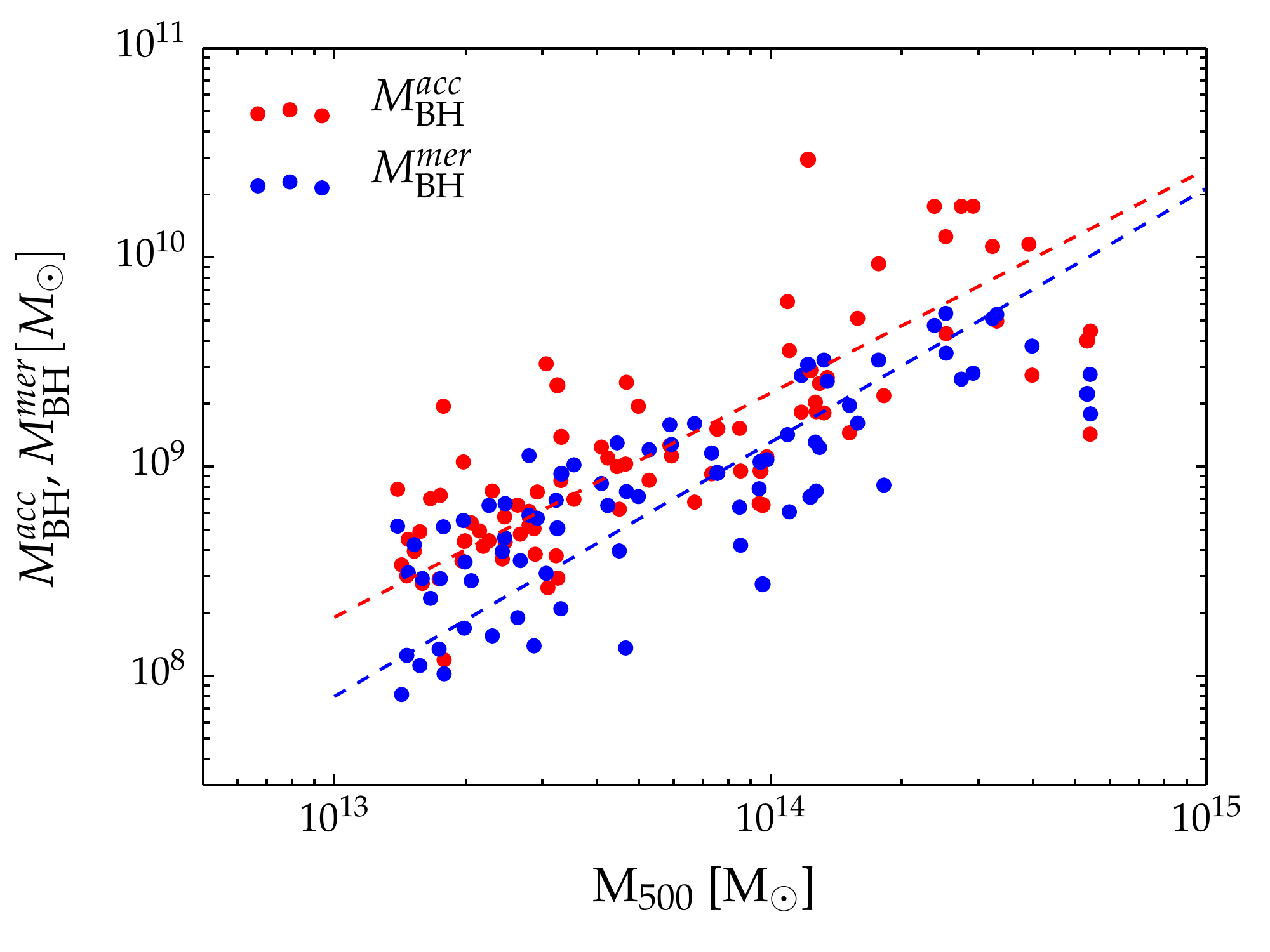}
	\includegraphics[width=\columnwidth]{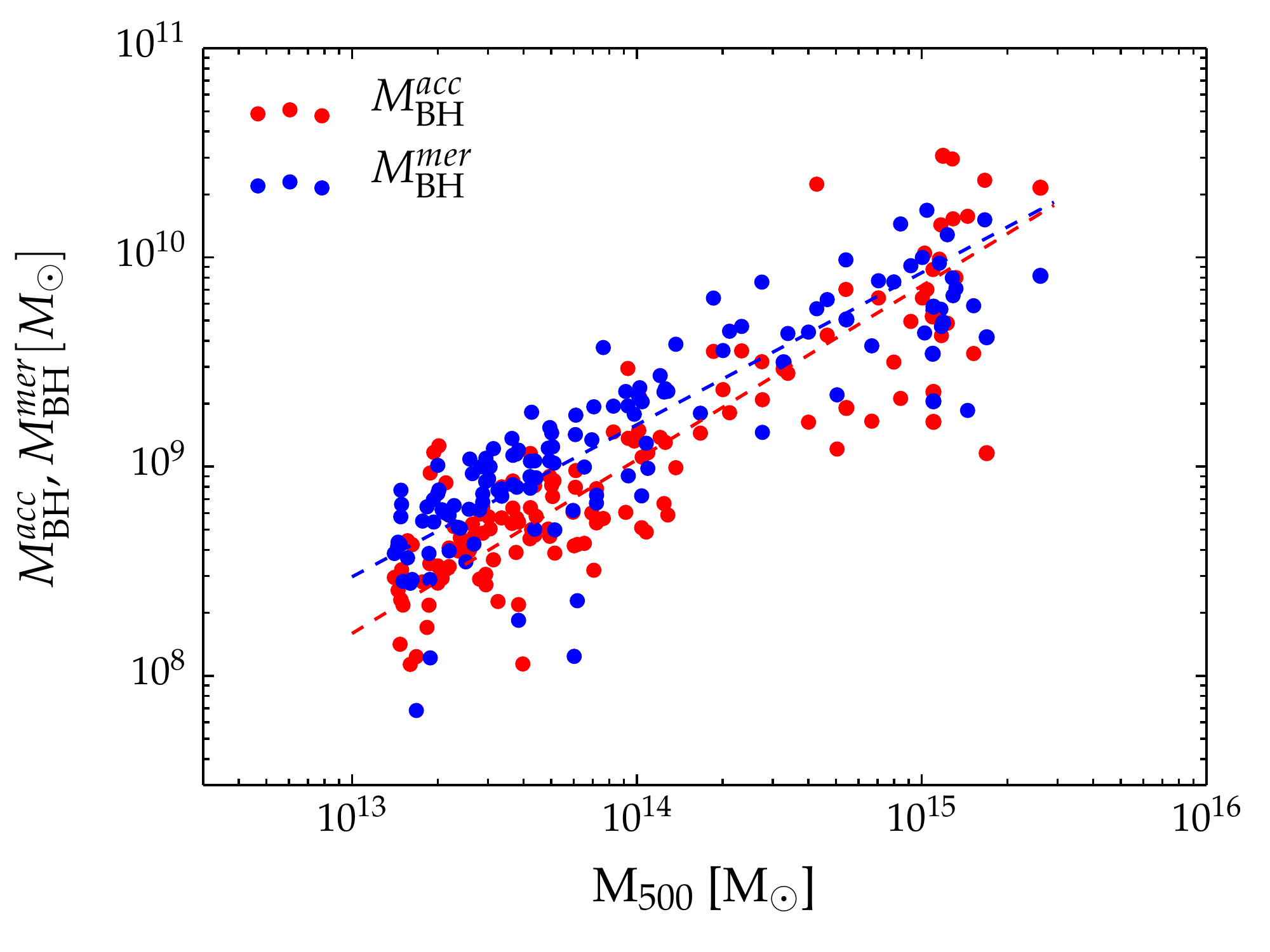}
    \caption{ Correlation between SMBH mass and M$_{500}$ at $z=1.0$ (left panel) and $z=0$ (right panel). We plotted in blue the mass gained by mergers and in red the mass gained by gas accretion. Dashed lines represent best fitting lines to the data.}
    \label{fig:m500_accretion_vs_merger}
\end{figure*}

\begin{figure}
	\includegraphics[width=\columnwidth]{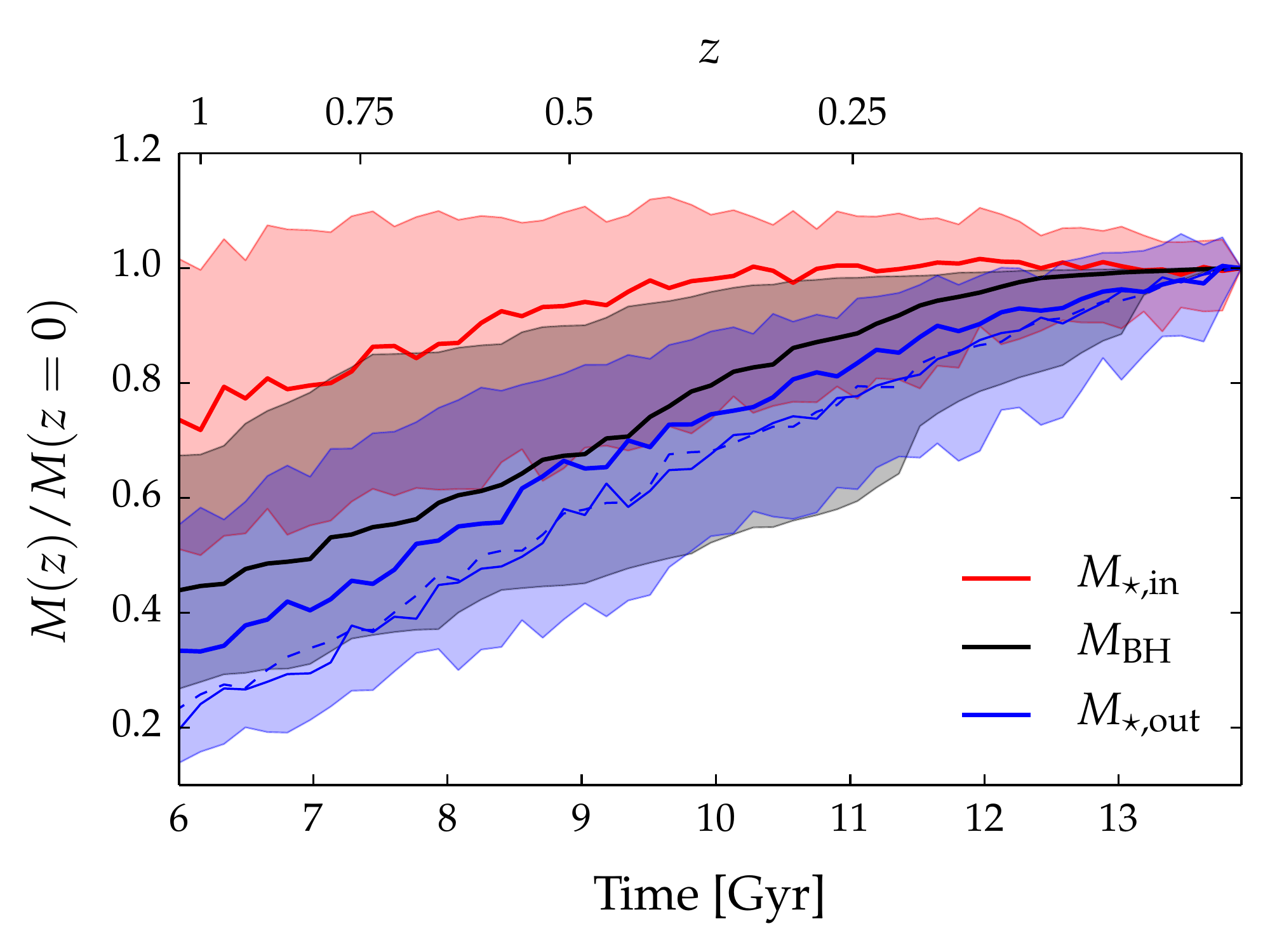}
    \caption{Evolution of $M_{\star, \rm in}$, defined as the stellar mass inside a spherical region of radius 30 kpc, SMBH mass, and $M_{\star, \rm out}$, defined as the total mass of the stars enclosed in a spherical shell with radii 30 kpc and 100 kpc. All the quantities are normalized to their respective values at $z=0$. Solid lines represent median values and shadowed regions represent 16 and 84 percentiles. The three different blue lines represent three definition of $M_{\star, \rm out}$. In particular the thin blue solid line is the stellar mass in a spherical shell with radii 100 kpc and 200 kpc while the dashed blue line is the stellar mass in a spherical shell with radii 50 kpc and 350 kpc.}
    \label{fig:icl}
\end{figure}

\subsection{Recent growth of SMBH and stellar component}

\noindent In our simulations the SMBH mass increases by a factor of $\sim 2.5$ between $1 > z > 0$. \cite{2018CINTHIA} found instead, selecting the most massive cluster in each Lagrangian region described in Sect.~2
, that the central stellar component measured within 30 kpc features a significant smaller growth. We checked that the growth factor of central SMBHs remain unchanged also considering this reduced subsample. This difference is due to the fact that in our simulations many substructures colliding with the BCG at $z < 1$ are largely disrupted and their stars quickly become part of the intra cluster light (ICL) or settle in the outermost radii of the BCG itself. This feature confirms some previous results. For example, \cite{2007MURANTE} shows that the bulk of the star component of the ICL originates during the assembly of the most massive galaxies in a cluster (and, in particular, of the BCG) after $z\sim 1$.

To visualize this effect, whose detailed study will require a dedicated analysis, we simply   compared the evolution of the SMBH mass and the inner stellar component, defined as $M_{\star, \rm in} = M_{\star} (r/{\rm kpc} < 30)$. Furthermore, we added a measure of the outer stellar component, defined as $M_{\star, \rm out} = M_{\star, \rm not-bound} (30  < r/{\rm kpc} < 100 )$. The "not-bound" identification specifies that we excluded all stars which are gravitationally bound to substructures identified by Subfind and different from the BCG. $M_{\star, \rm out}$, therefore, comprises both the ICL and the outermost stellar mass of the largest BCG in the sample. The median values of  $\Delta M_{\star, \rm in}$, $\Delta M_{\rm BH}$ and $\Delta M_{\star,\rm out}$ are computed after the normalization to their respective mass values at $z=0$. 

We considered 30 kpc for $M_{\star, {\rm in}}$ because we wanted to evaluate the changes on the stellar component in the immediate surrounding of the SMBH; furthermore, this was used in \cite{2018CINTHIA} as one of the possible definition of the BCG mass. For the outer component, instead, we also considered the mass of the unbound stars measured in other two spherical shells: between 100 kpc and 200 kpc and between 50 kpc and 350 kpc. As clear from Fig.~12, the choice of this region does not substantially change our conclusions: while $M_{\star, \rm in}$ slowly increases from $z=1$ to $z=0$, the SMBH mass and $M_{\star,\rm out}$ rapidly increase. At $z=1$, indeed, the quantities are about 80 percent, 45, and 35 per cent of their final values, respectively. The remarkable agreement between the two extra definitions of the outer stellar component -- $M_{\star, {\rm not-bound}} (100  < r/{\rm kpc} < 200$) and $M_{\star, {\rm not-bound}} (50  < r/{\rm kpc} < 350$), thin blue solid and dashed line in Fig.~12 -- implies that the growth rate of the ICL is independent on the specific radius used. 

The final emerging picture is that many small substructures actually reach the cluster core and merge with the BCG. However, few of their stars remain in the innermost region. The interaction with the BCG causes that most stars of the structures are tidal shocked and stripped. Subsequently, they become gravitationally unbound thereby taking part of the ICL. During the disruption of the substructures, their most massive SMBHs feel the gravitational attraction of the underlying potential and sink towards the minimum of the cluster potential contributing to the growth of the SMBH at the center of the BCG. We note, however, that the modeling of BH-BH mergers is very simplistic and could overestimate the efficiency of this physical process which take place at a scale well below the gravitational softening of the simulations.

\subsection{The $M_{\rm BH}-M_{500}$ relation for the two SMBH growth channels }

Given that at $z=1$ and $z=0$ the SMBHs have grown from both channels (through gas accretion and BH-BH mergers), it is relevant to check whether the SMBH mass of the two channels are separately both related to the total mass or whether only one exhibits a tight correlation while the other mostly contributes to increase the scatter. This possibility is investigated in Fig. \ref{fig:m500_accretion_vs_merger} where we considered both $M_{\rm BH}^{\rm acc}$ and $M_{\rm BH}^{\rm mer}$ as a function of M$_{500}$ at $z=1$ (left panel) and at $z=0$ (right panel). As we have seen, at $z=1$ the gas accretion is the dominant channel. In Sect.~4.4, we saw that this channel  shows only a slight increment from $z=1$ to $z=0$. For this reason, the red points, referring to $M_{\rm BH}^{\rm acc}$ are substantially unchanged in the two panels. Viceversa, $M_{\rm BH}^{\rm mer}$ becomes the dominant component at $z=0$. The results of the linear fits of the relations are reported in Table~\ref{tab:all} also for the other redshifts.

From the figure, it is evident that both masses correlate well with M$_{500}$ at both redshifts independent of which one of the two is the dominant channel from the SMBH mass growth. From the table, we notice that the slopes of the two relations are consistent within 1 $\sigma$ being the $z=1$ slightly steeper as expected from the Sect.~4.3. Most importantly, the two scatters are similar and both slightly higher than the scatter of the relation of the total SMBH mass (see Table~\ref{tab:all}). 

Finally, we would like to emphasize that Fig.~7 suggests that the $M_{\rm BH}-M_{500}$ relation is already in place at $z=2$ when the SMBH mass was almost entirely gained only by gas accretion. These results enlighten that, at least in our simulations, mergers are not essential to establish the relation at first. 

\section{Discussion}

As previously remarked the correlation between the BCG mass and the SMBH mass has been diffusely studied and often used to extract the mass of the SMBH knowing the mass of the hosting galaxy (Sect.~1). In \cite{2018BOGDAN}, the authors found in their observed sample that the scatter between the SMBH mass and the global cluster properties is tighter by almost 40 per cent than the scatter of the $M_{\rm BH}-M_{\rm BCG}$ relation. Indeed, in their Table~4 they report $\sigma_{M_{\rm BH}|M_{\rm BCG}}=0.61$ and $\sigma_{M_{\rm BH}|T}=0.38$. The values of the scatters increase in the later analysis by \cite{2019PHIPPS} because of the method used to derive $M_{\rm BH}$, the authors suggest. Despite, also in that case the two intrinsic scatters in $M_{\rm BH}$ are comparable between each other. As a consequence, the global properties of cluster within $R_{500}$ are also suitable to estimate the SMBH mass.
We show in Sect.~4 that the distribution of our simulated data is in reasonable agreement with the observations by \cite{2018BOGDAN} and \cite{2019GASPARI}, who used dynamical measurements of $M_{\rm BH}$. We  demonstrate that for our simulated objects there is a clear correlation between the mass of the SMBHs, located at the minimum of the potential well, and the temperature or total mass of the clusters within $R_{500}$. In this section we discuss the properties of all the relations described in the paper and listed in Table~2.

\begin{figure*}
	\includegraphics[width=2\columnwidth]{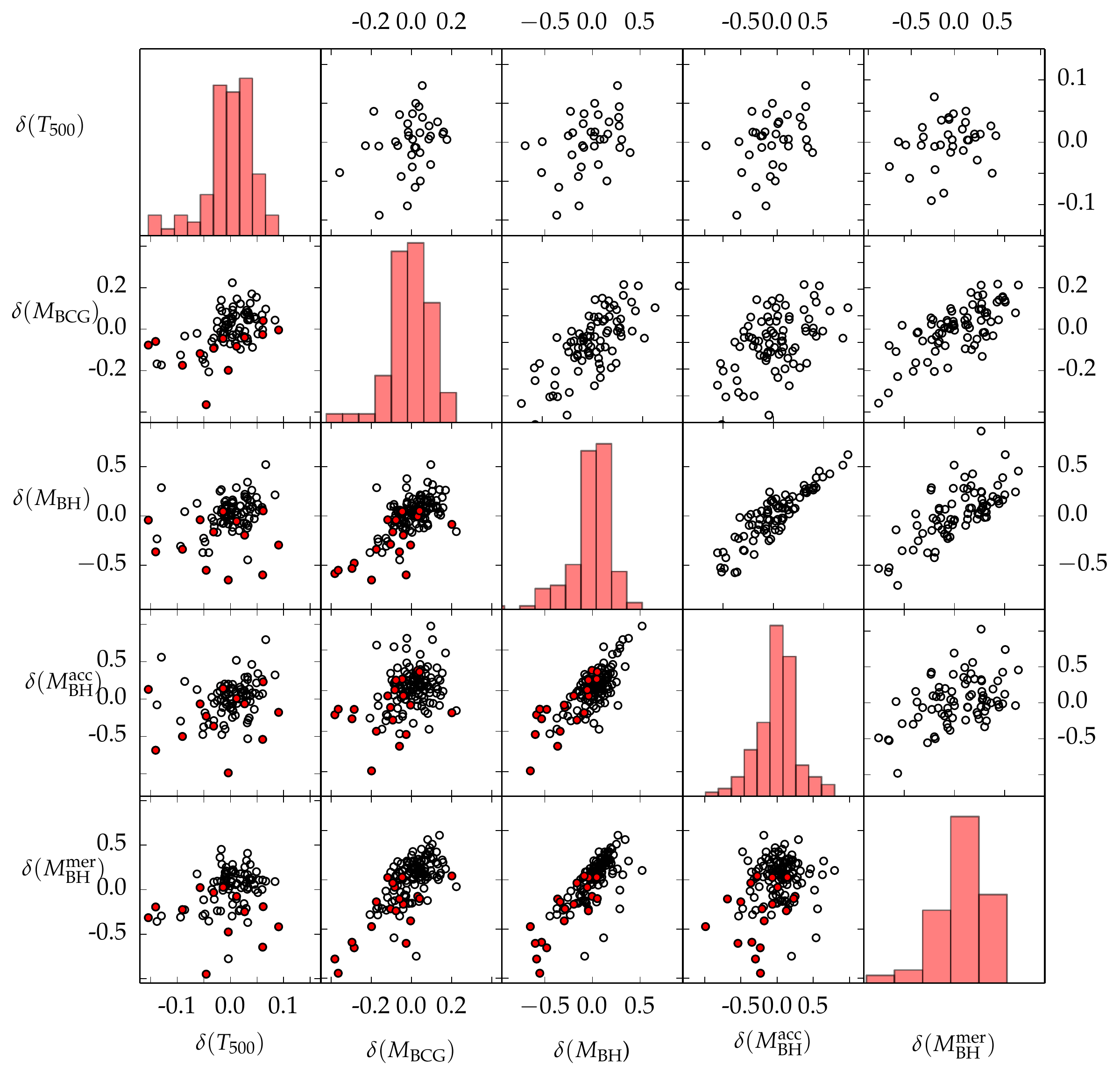}
    \caption{Covariance matrix. Each axis represents the logarithmic difference between the actual value of a quantity X and the expected value from the linear relation (X-$M_{500}$) at its fixed mass, as defined by Eq.~\ref{eq:delta}. Panels above the diagonal refer to $z=1$ while panels below the diagonal refer to $z=0$. The diagonal panels show the distribution of $\delta(X)$ at $z=0$. Red points define clusters which have experienced a mass growth of at least $40 \%$ during the last Gyr. }
    \label{fig:total_plot}
\end{figure*}

To this end we refer to Fig. \ref{fig:total_plot} where we show the covariance matrix between the deviations of all the quantities of interest from their best-fitting relations, that are their residual at fixed mass. These are defined as logarithmic differences between the actual value, generically referred as $X$, and the expected value, $X_{\rm FIT}$, from the $X-M_{500}$ relations\footnote{The $(M_{500},T_X)$ relation has previously been inverted into $(T_X,M_{500})$.} provided in Table~2:
\begin{equation}
\delta(X)=\log[X/X_{\rm FIT}].
\label{eq:delta}
\end{equation}
The deviations are computed for each quantity $X=T_{500}$, $M_{\rm BCG}$, ${M_{\rm BH}}$, ${M_{\rm BH}^{\rm mer}}$, ${M_{\rm BH}^{\rm acc}}$. The panels above the diagonal refer to $z=1$ while those below to $z=0$. The diagonal panels show the distribution of $\delta(X)$ at $z=0$. We used  the temperature subsample whenever $\delta(T_{500})$ is considered. For each pair of deviations we listed the Spearman correlation coefficients in Table~3 computed at $z=0$, $0.5$, $1$ and we marked in bold the correlations whose probability to be consistent with zero is less than 2 per thousand and its module is greater than 0.4. A strong correlation between $\delta(M_{\rm BH})$ and $\delta(X)$ is converted into a small scatter in the relation $M_{\rm BH}-X$.

\begin{table*}
\caption{Correlation matrix. Each element represents the Spearman coefficient of the quantities of interest. The three values refer to $z=0$, $z=0.5$ and $z=1$ respectively. The values in bold have probability to be different from 0 below 0.2 per cent.}
\begin{tabular}{c|cccc}
&$\delta (T_{500})$ &$\delta (M_{\rm BCG})$ &$\delta (M_{\rm BH})$ &$\delta (M_{\rm BH}^{\rm acc})$ \\
\hline
$\delta (M_{\rm BCG})$       &   {\bf  0.498}|  0.072 | 0.255         &       &  &   \\
$\delta (M_{\rm BH})$          &   0.322   |  0.210  | 0.355         &{\bf  0.557} | { \bf 0.693} | { \bf 0.694}        &  & \\
$\delta (M_{\rm BH}^{\rm acc})$&   0.310   |     0.243  | 0.267     & 0.288 | {\bf  0.405} | { \bf 0.485} &{\bf  0.684} | {\bf  0.804} | { \bf 0.865}& \\
$\delta (M_{\rm BH}^{\rm mer})$ &  0.230   | 0.087 |  0.277 &  {\bf  0.612} | {\bf  0.751} | {\bf  0.679} &{\bf  0.777} | {\bf  0.753} | {\bf  0.737} & 0.190 | {\bf 0.304} | {\bf 0.384}       \\
\end{tabular}
\end{table*}

 As previously commented, the scatter $\sigma_{M_{\rm BH}|T_{500}}$ is comparable to $\sigma_{M_{\rm BH}|M_{500}}$ (see Table~2). As a consequence, the temperature and the total cluster mass are equally good proxy for the SMBH mass. The similarity between these two proxies can be explained by looking at the correlation between $\delta(T_{500})$ and $\delta(M_{\rm BH})$ in Fig.~13. The panels for the two considered redshifts, $z=0$ and $z=1$, highlight how the variations of cluster temperature are not directly reflected into variations of the SMBH mass. At first sight, this can be surprising as one could expect that both quantities are strongly dependent on the dynamical activities of the cluster core and, especially, sensitive to merger events that impact the innermost region of the clusters. However, the response of the temperature and the SMBH mass to merger is not simultaneous: we saw in Sect 4.2 that the typical delay between the increase of the $M_{\rm BH}$ after a merger can be around 1-3 Gyr, while the temperature increase typically occurs in less than 1 Gyr from the merging episode. In addition, the rapid increase of the ICM temperature is followed by a small drop caused by the expansion of the shock towards the more external regions. This temperature oscillation along with the mismatch between the two time-delays explain why the correlation coefficients between $\delta(T_{500})$ and all $M_{\rm BH}$ components are always low and with a high probability to be consistent with zero. These characteristics are also in place when the temperature variations are compared with the variation on the SMBH mass due to BH-BH mergers. 

When comparing, instead, the scatter of the relation between the SMBH mass and the cluster properties with the scatter of the $M_{\rm BH}-M_{\rm BCG}$ relation, we find that $\sigma_{M_{\rm BH}|M_{\rm BCG}}$ is smaller at all times. At redshift $z=0$, for example, $\sigma_{M_{\rm BH}|M_{500}}$ is $\approx 25$ per cent larger, although the two scatters are in agreement between $2 \sigma$. The difference between the scatters is more significant at $z\ge 1$ where typically $\sigma_{M_{\rm BH}|M_{500}}$ is $\approx 1.4 \times \sigma_{M_{\rm BH}|M_{\rm BCG}}$. In the covariance matrix formalism this translates into a correlation between $\delta(M_{\rm BCG})$ and $\delta(M_{\rm BH})$. Indeed, at $z \leq 1$ the correlation is $\geq$ 0.55. Moreover, the correlation is stronger if computed with respect to $M_{\rm BH}^{\rm mer}$ when it reaches values around 0.7. As we can see from Fig.~13, the correlation at $z=0$ is significant for the presence of situations of either pre-mergers or mergers with a large impact parameter when both $M_{\rm BCG}$ and $M_{\rm BH}$ (and in particular $M_{\rm BH}^{\rm mer}$) are smaller than the average of the sample at fixed total mass (e.g., their $\delta$ is negative). At $z=1$, we also notice a correlation between the variations of the two quantities in the other direction (both $\delta(M_{\rm BCG})$ and $\delta(M_{\rm BH}^{\rm mer})$ greater than zero) implying that the SMBHs that gained mass through $z\geq1$ mergers are hosted in BCG with higher stellar mass with respect to the average of the sample. 

In Fig.~5, we show that the scatter of the $M_{\rm BH}-M_{500}$ relation is influenced by the presence of the systems that recently experienced a major merger. Indeed, all objects that increased their total mass by at least 40 per cent in the last Gyr are in the bottom part of the overall distribution (see Sect.~4.2 and Fig.~5). In Fig.~13 we identify these objects as red points. In the majority of the cases their variations with respect to the mean are negative. When we exclude these objects and reapply our fitting procedure\footnote{The best-fitting relations of the samples derived by excluding the clusters with recent fast accretion are: \\
$M_{\rm BH}/(10^9 {\rm M_{\odot}}) = 10^{0.49 \pm 0.02} \times (M_{500}/10^{14} {\rm M_{\odot}})^{0.77 \pm 0.03}$; \\ 
$M_{\rm BH}/(10^9 {\rm M_{\odot}}) = 10^{0.56 \pm 0.02} \times (T/2 {\rm keV})^{1.32 \pm 0.06}$; \\ 
$M_{\rm BH}/(10^9 {\rm M_{\odot}}) = 10^{-0.41 \pm 0.03} \times (M_{\rm BCG}/10^{11} {\rm M_{\odot}})^{1.17 \pm 0.04}$.},
we find that the scatter of the $M_{\rm BH}-M_{500}$ relation reduces by almost 20 per cent ($\sigma_{M_{\rm BH}|M_{500}}=0.14 \pm 0.01$), it remains comparable to the re-computed scatter of the $M_{\rm BH}-T_{500}$ relation ($\sigma_{M_{\rm BH}|T_{500}}=0.14 \pm 0.02$) and consistent with the new $\sigma_{M_{\rm BH}|M_{\rm BCG}} = 0.13 \pm 0.01$. In other words, even if the removal of the dynamically active objects induces a decrease in the scatters of all of the three relations, the most important reduction impacts the scatter at fixed total mass, reducing the gap between $\sigma_{M_{\rm BH}|M_{\rm BCG}}$ and the scatter of the relations involving global cluster properties. 

In the interpretation of these results from simulations, we need to recall that the simulated data do not reproduce the observed scatter of the $M_{\rm BH}-M_{\rm BCG}$ relation (Fig.~1 and also \citealt{2013RAGONE, 2018BOGDAN}). On one hand the growth of simulated SMBHs is regulated by simplistic subgrid models that do not capture all physical processes in place and might lead to a reduced scatter. On the other hand, as explained when discussing Fig.~1, a large portion of the observed scatter around the $M_{\rm BH}-M_{\rm BCG}$ relation can be ascribed to observational uncertainties associated either with the quantity definition (e.g., treatment of intra-cluster light, BCG boundary definition) or with the measurement procedures (e.g., not fixed aperture mass for the BCG or application of scaling relation to infer the SMBH mass). In simulations, instead, the BCG and SMBH masses are always known and precisely defined (see Sect. 2.1 on the discussion on the BCG mass determination). 
%
%
These arguments not only provide a possible explanation for the difference between the simulated and observed scatters but also underline that the errors on the measures of $M_{\rm BCG}$ derived from observations are not easily reducible. The estimate of the SMBH mass from the BCG mass can always be subject to these uncertainties. The global cluster properties are also subject to systematics, which however can be treated as follows. A systematic bias on the global temperature can be dealt with precise instrument calibration or with multi-temperature fitting. The uncertainties on the total mass are reduced when measurements coming from various wavelengths are combined, such as mass reconstruction from gravitational strong and weak lensing, galaxy dynamics, SZ, and X-ray. These considerations, along with the limited difference in the relation scatters, emphasize how the global cluster properties can be powerful proxies for the SMBH mass. This conclusion is even stronger at high redshift, such as $z=1$.


\section{Conclusion}

In this paper we studied the correlation between the mass of SMBH at the center of BCG and global properties of the hosting cluster of galaxies, namely its total mass, $M_{500}$, and global temperature, $T_{500}$. Our work is based on 29 zoom-in cosmological hydrodynamical simulations carried out with GADGET-3, a modified version of the public code GADGET-2. This code treats unresolved baryonic physics including AGN feedback through various sub-grid models. The parameters used to model the AGN feedback are tuned to appropriately reproduce the scaling relation between the masses of the SMBHs and their hosting galaxy (Fig.~1).   
For this study, we considered all systems with $M_{500} > 1.4 \times 10^{13}$ M$_{\odot}$ identified in the high-resolution regions of the re-simulated volumes. At $z=0$, there are 135 objects. 

After showing the agreement between our numerical results and the observational data, we explored how the relation between the SMBH mass and the cluster mass establishes by looking at the co-evolution of these quantities in individual systems. We then looked at the evolution of the entire sample considering four different times ($z= 0.5, 1, 1.5$, and $2$). Finally, we characterized the role played by the two channels for the SMBH growth: accretion of gas and BH-BH merging. Our main results can be summarized as follows.

\begin{itemize}

    \item[$\bullet$] The simulated relations between SMBH mass and the global cluster properties ($M_{\rm BH}$-$M_{500}$ and $M_{\rm BH}-T_{500}$) are in agreement with the observations by \cite{2018BOGDAN} and \cite{2019GASPARI} (Fig.~2 and Fig.~4).
    
    \item[$\bullet$] The $M_{\rm BH}-M_{500}$ relation at $z=0$ originates from a non-simultaneous growth of the SMBH and of the cluster. In particular, objects evolve on the $M_{\rm BH}-M_{500}$ plane either at almost constant $M_{\rm BH}$ or at almost constant $M_{500}$ (Fig.~5). The rapid increase of the cluster total mass occurs during cluster mergers and typically last 1 Gyr. Subsequently, the substructures move towards the cluster center and, eventually, reach the core feeding the central SMBH with gas and/or inducing a BH-BH merger. Clusters that recently experienced major merger events are in general below the mean relation.
    
    \item[$\bullet$] The $M_{\rm BH}-M_{500}$ relation of the entire sample shows a degree of evolution: the slope is about 45 per cent smaller at $z=0$ with respect to $z=2$ (Fig.~7). This evolution naturally arises from hierarchical structure growth as the most massive clusters increase their mass in the period between $z=2$ and $0$ at a rate which is 10-20 times higher with respect to the smallest objects. Viceversa, the SMBH mass increases by an approximately constant factor ($\approx 5$), independent of the mass of the hosting cluster (Fig.~6).
    
    \item[$\bullet$] In our simulations, SMBHs grow by two different channels. Gas accretion is the most relevant channel at redshift $z>2$ and the only player at the earliest times. The accretion is slowed down only when the SMBHs are massive enough to balance gas cooling via AGN feedback. At lower redshift ($z=1$) one quarter of the SMBH mass is ascribed to mergers. From that time to $z=0$ the BH-BH merger contribution becomes progressively more important. Indeed, mergers contribute by about 60 percent of the total $z=0$ SMBH mass on average. When restricting the analysis to the most massive systems, we find that the accretion onto the SMBH is dominant up to $z=0$, both for a reduced power of the AGN heating over the gas cooling and for a less frequent BH-BH merger rate.    
    $M_{\rm BH}^{\rm acc}$ and $M_{\rm BH}^{\rm mer}$ similarly relate to $M_{500}$ at both $z=0$ and $z=1$, independently of which SMBH mass growth channel is dominant.
    
    \item[$\bullet$] The rapid increase of the SMBH mass at $z < 1$ due to mergers is much faster than that of the hosting BCG, defined as the stellar mass within the fixed physical aperture of either 30 or 50 kpc and studied in details by \cite{2018CINTHIA}. Indeed, in our simulations, we found that at recent times ($z<1$) galaxies are frequently merging with the BCG but they get easily disrupted. This process rather than transferring the merging stellar mass to the BCG originates and feeds the diffuse stellar component (the intra-cluster light), that we simply define here as unbound stars located outside the BCG.     
    
    \item[$\bullet$] The $M_{\rm BH}-M_{500}$ and $M_{\rm BH}-T_{500}$ relations present a similar scatter (or, equivalently, $\delta(T_{500})$ does not show any significant correlation with $\delta(M_{\rm BH})$), meaning that they are equally valid SMBH mass proxy. On the other hand, $\delta(M_{\rm BH})$ is highly correlated with $\delta(M_{\rm BCG})$. As a consequence the scatter of the $M_{\rm BH}-M_{\rm BCG}$ relation, at $z=0$, is $\approx 25 \%$ lower in our simulations, although the two values are in agreement within $2 \sigma$. However, it is important to stress that the observed scatter of the $M_{\rm BH}-M_{\rm BCG}$ relation is larger then the simulated one mainly for two reasons: the numerical limitation of a simplistic description of SMBHs growth and the large uncertainties affecting the observational measurements of both SMBH and BCG mass. That said, when the most dynamically active objects are discarded from the sample, the scatters of all relations, $M_{\rm BH}-M_{\rm BCG}$, $M_{\rm BH}-M_{500}$ and $M_{\rm BH}-T_{500}$, become similar, thus strengthening the predicting power of the cluster global quantities. 
    

\end{itemize}

Even though our simplified sub-grid models are effective in reproducing the observed relations, before concluding, it is important to list some limitations of our current configuration, which we aim at improving in future work. As first point our resolution is so-far limited to about 5 kpc to obtain a large statistical sample in a cosmological environment. This means that while the large-scale inflows can be correctly described, satellite galaxies are resolved with few smoothing lengths implying that the actual level and timing of gas accretion and SMBH mergers might differ from our tracked evolution. 
Resolving satellite galaxies with less than $\sim 250$ particles results in reducing their concentration and ability to resist to disruption events \citep{2018MNRAS.475.4066V}. This might be one of the causes of a slightly reduced number of galaxies with $M_{\star}\sim 10^{11}\ M_{\odot}$ in our simulations with respect to the field stellar mass function as measured by \cite{2013MNRAS.436..697B} (see Fig.~\ref{fig17}). Similar problems affect other simulations \citep{2015MNRAS.448.1504S} and have been discussed in connection to the mismatch between the substructure distribution measured in observations and simulations (e.g., \citealt{2015ApJ...800...38G}; \citealt{2016ApJ...827L...5M}). Explanations for the discrepancy between the observed and simulated stellar mass functions can also be of physical nature rather than exclusively numerical. For example, a more physically motivated parametrization of the AGN feedback could play a relevant role. 

Several studies (\citealt{2016GASPARI} for a brief review) show that the SMBH accretion in massive galaxies proceeds via chaotic cold accretion (CCA), from the kiloparsec scale down to tens gravitational radii.   While our subgrid feeding module tends to mimic CCA in power, the rapid frequency of chaotic events is not fully captured here. CCA leads to the rapid funneling of cold clouds also at low redshift, as the hot halo develops nonlinear thermal instability and a consequent rain of clouds toward the inner SMBH. Such frequent accretion, which is key to quench cooling flows, would likely extend the dominance of gas feeding over mergers even at low $z$. Our injection of AGN feedback is also simplistic, since we only inject thermal energy within a fixed aperture with a fixed efficiency. The physics of feedback is quite more complex: massive outflows and radio jets inflate bubbles and drive shocks in the ICM, while CCA seems to occur in perpendicular cones (e.g., \citealt{2018GASPARI}; \citealt{2019TREMBLAY}). This leads to highly variable duty cycle and variations in the core ICM properties (temperature, entropy, etc.). Finally, we can improve the predictive power of future simulations by using physically motivated parameters, such as the horizon mechanical efficiency given in GR-MHD BH accretion simulations (e.g., \citealt{2017GASPARI}).
    

\begin{acknowledgements}
We thank the anonymous referee for the careful and constructive reading of the paper and for his/her useful suggestions. We would like to thank Marta Volonteri and Lorenzo Lovisari for helpful feedback during the draft of the paper and Volker Springel for making the GADGET-3 code available to us.
This project has received funding from: ExaNeSt and Euro Exa projects, funded by the European Union’s Horizon 2020 research and innovation program under grant agreement No 671553 and No 754337, the agreement ASI-INAF n.2017-14-H.0, 
the Agencia Nacional de Promoci\'on Cient\'ifica y Tecnol\'ogica de la Rep\'ublica Argentina under the PICT2417-2013 grant; the Consejo Nacional de Investigaciones Cient\'ificas y T\'ecnicas de la Rep\'ublica Argentina (CONICET); the Secretar\'ia de Ciencia y T\'ecnica de la Universidad Nacional de C\'ordoba - Argentina (SeCyT); the European Union’s Horizon 2020 Research and Innovation Programme under the Marie Sklodowska-Curie grant agreement No 734374, PRIN-MIUR 2015W7KAWC, the INFN INDARK grant, NASA Chandra GO7-18121X {  and GO8-19104X}. 
Simulations have been carried out using MENDIETA Cluster from CCAD-UNC, which is part of SNCAD-MinCyT (Argentina) and MARCONI at CINECA (Italy), with CPU time assigned through grants ISCRA C, and through INAF-CINECA and University of Trieste - CINECA agreements. The post-processing has been performed using the PICO HPC cluster at CINECA through our expression of interest. 
\end{acknowledgements}

\bibliographystyle{aa}
\bibliography{biblio}

\begin{appendix}

\section{Effects of feedback parameters on SMBH growth}

Cosmological hydrodynamical simulations rely on a number of subgrid models employed to simulate physical processes not resolved by the simulation itself (e.g., star formation and stellar feedback, chemical evolution, and AGN feedback). These models depend upon a set of parameters, tuned in order to reproduce some observational constraints. In particular in our simulations the parameters regulating the AGN feedback are chosen in order to reproduce the observed Magorrian relation \citep{MAGORRIAN1998}. However, \cite{2016SHANKAR} and \cite{2019SHANKAR} have suggested that the observed correlation between stellar mass and SMBH mass might be biased high, because the SMBHs sphere of influence needs to be resolved in order to measure SMBHs mass with spatially resolved kinematics. 

We, therefore, verified whether our results are influenced by a change of the normalization on the Magorrian relation and in which direction. In our sub-grid model the normalization is regulated only by the radiative efficiency value, $\epsilon_f$, that is the fraction of the energy radiated by AGN feedback which is thermally coupled to the surrounding gas particles. To perform this test, we used two simulations of the same Lagrangian region centered around a cluster with $M_{500}=3.38\times 10^{14}\ M_{\odot}$ at a mass resolution 10 times higher than the one used for the simulations presented in the paper and therefore also more spatially resolved. The two simulations differ only for the feedback efficiency $\epsilon_f$ of the quasar-mode set equal to $0.1$ in one simulation and to $0.15$ in the other, while in the radio mode the parameter is kept equal to 0.7.

\begin{figure}
	\includegraphics[width=\columnwidth]{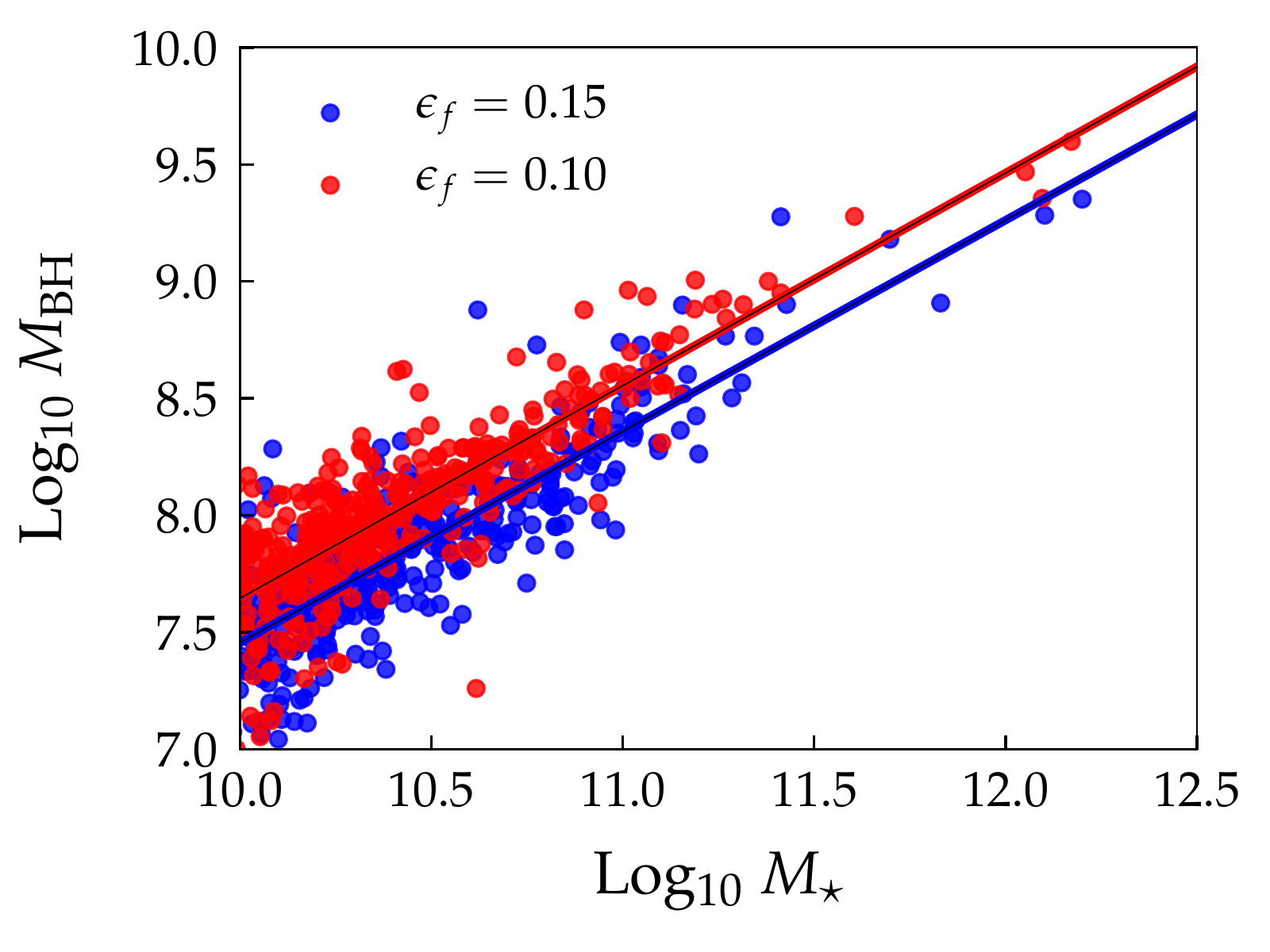}
    \caption{Correlation between SMBH mass and stellar mass for the two different values of the radiative efficiency $\epsilon_f$. Blue and red solid lines represent best fit to the data.}
    \label{fig:magorrian_fb}
\end{figure}

In Fig. \ref{fig:magorrian_fb} we show the effect of the feedback efficiency on the $M_{\rm BH}-M_{\star}$ relation. As we commented before the only effect is a change in the normalization, while the slope and the scatter are similar between the two runs. This result has been already studied both from a numerical (e.g., \citealt{2009BOOTH}) and theoretical (e.g., \citealt{2005CHURAZOV}) point of view, therefore is not totally unexpected. 

\begin{figure}
	\includegraphics[width=\columnwidth]{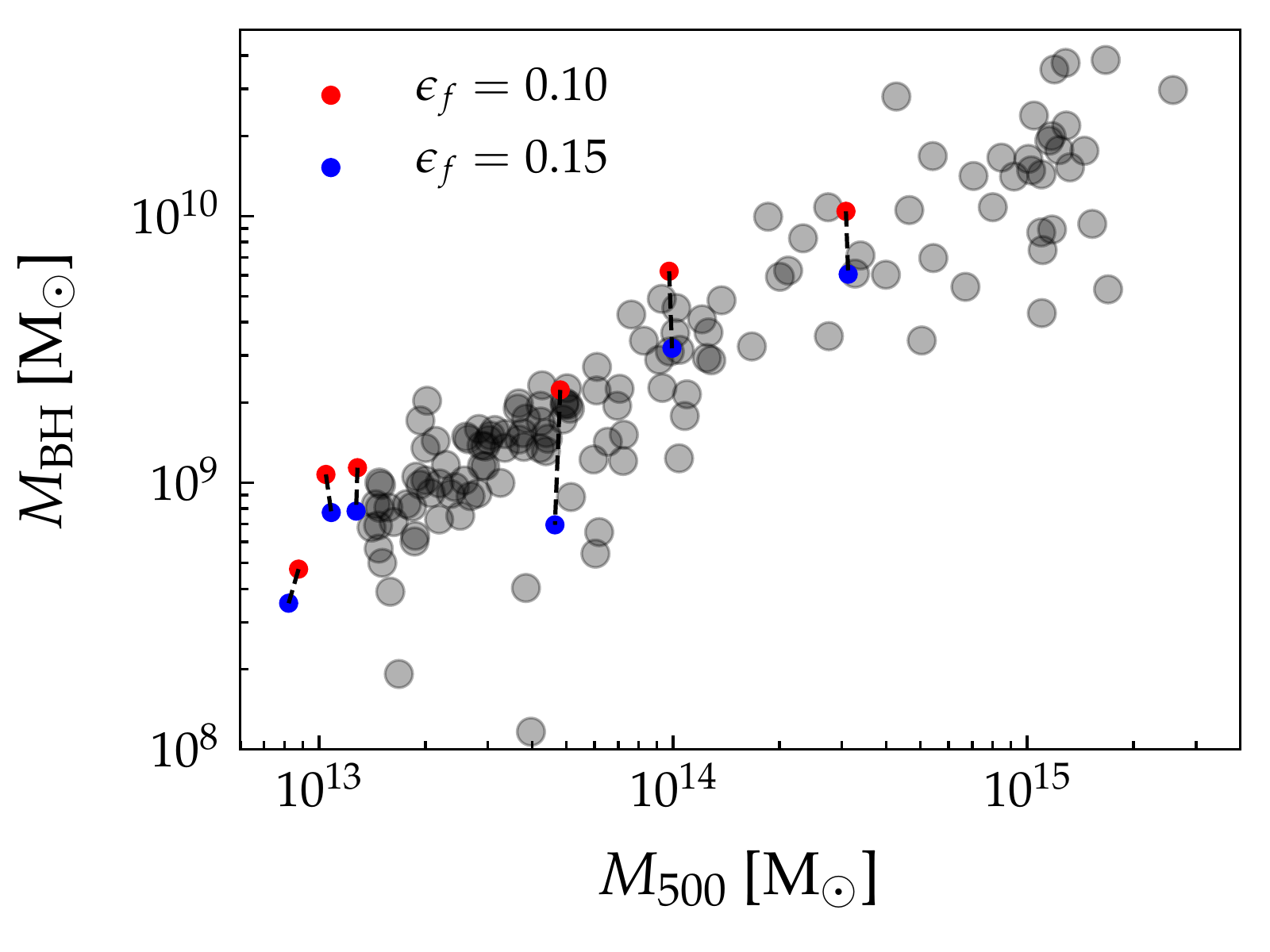}
    \caption{Correlation between SMBH mass and cluster mass. Black points represents simulated data of Fig.~4. Red and Blue points represent results of the two simulations of the same region at higher resolution. Dashed lines connect the same systems in the two simulations.}
    \label{fig:fb1}
\end{figure}

 As second check, we studied the effect on the $M_{\rm BH}-M_{500}$ relation presented in Sect.~4.1. From Fig\ref{fig:fb1} we see that also in this case a higher feedback efficiency leads to less massive black holes, and thus to a lower normalization of the $M_{\rm BH}-M_{500}$ relation. There is no effect on the total mass of the clusters. Indeed, $M_{500}$ varies by little but not in a systematic direction: it might slightly increase as decrease. 

Finally, we study what is the effect of changing $\epsilon_f$ in the relative importance of the SMBH growth channels presented in Sect.~4.4: $M_{\rm BH}^{\rm mer}$ and $M_{\rm BH}^{\rm acc}$. 
\begin{figure}
	\includegraphics[width=\columnwidth]{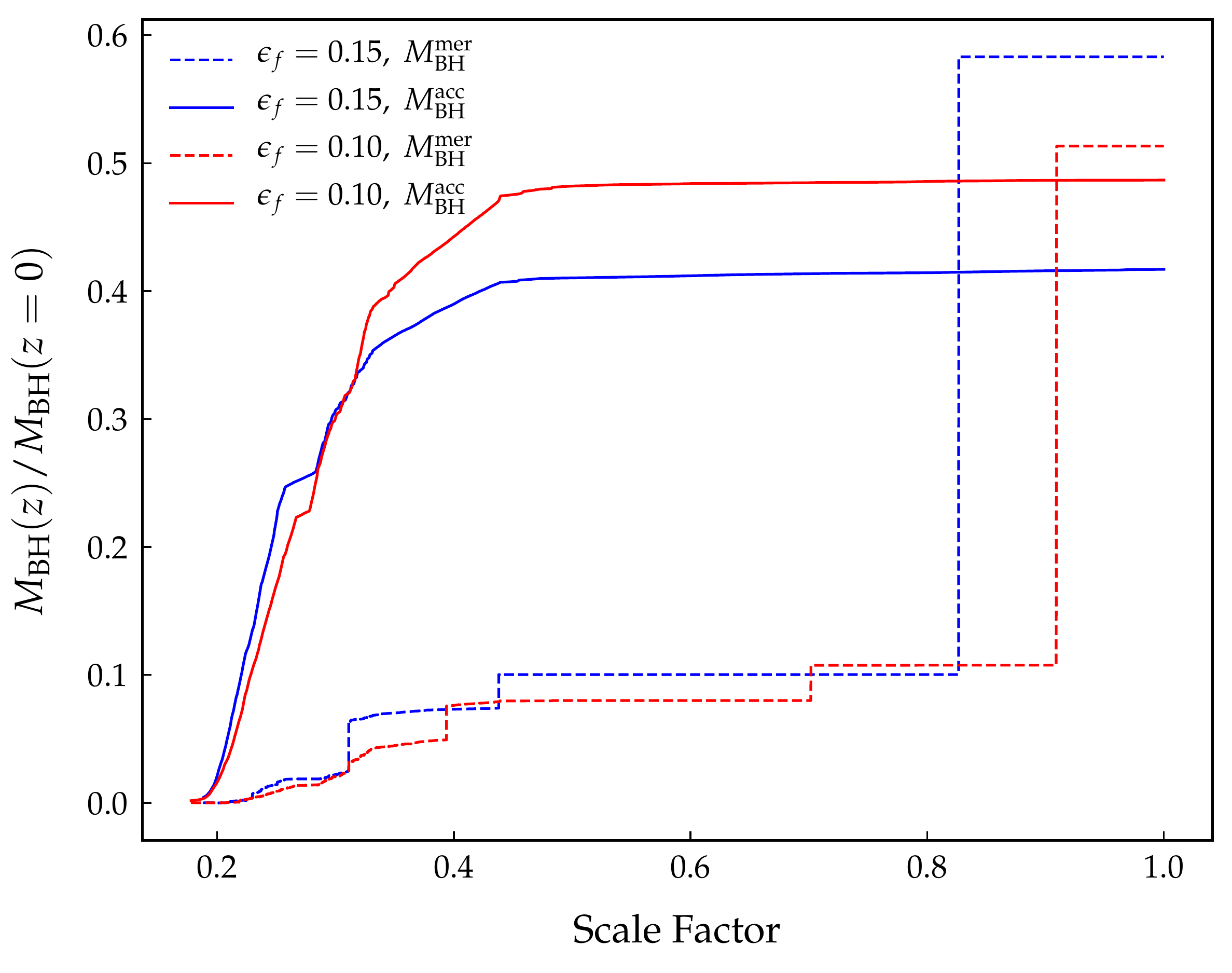}
    \caption{Evolution of the mass of the same SMBH in the two simulations at high resolution. The solid lines represent the growth of the mass of the SMBHs due to gas accretion. Dashed lines represent the mass gained via BH-BH mergers.}
    \label{fig:fb2}
\end{figure}
In Fig.~\ref{fig:fb2} we show the equivalent of Fig.~8 for the SMBH at the center of the main cluster in the two simulations. We see that a lower value of $\epsilon_f$ (red lines) leads to a fraction of mass due to gas accretion which is higher by about 20 percent. We repeated the same analysis on six SMBHs that we could identify in the Lagrangian region and match as part of corresponding clusters. We find that the median of the ratios between the fraction of SMBH mass due to gas accretion in the two simulations is 0.83, confirming that the plot is representative of the difference in the BH growth.

 In conclusion, if we chose a larger value of $\epsilon_f$ to match a lower normalization of the $M_{\rm BH}-M_{\star}$ relation as suggested by \cite{2016SHANKAR}, we expect that the $M_{\rm BH}$ will decrease but both the stellar mass and the cluster mass remains unchanged. As a consequence the scatter and slope of the $M_{\rm BH} -M_{\star}$ and the $M_{\rm BH}-M_{500}$ relations will maintain their values. Finally, the limited growth of the SMBH is mostly caused by a much reduced level of gas accretion which is not fully replenished by the BH-BH mergers. In fact, even thought the merger events play a more significant role with respect to the accretion, all the merged SMBHs are on their own smaller for the less effective accretion experienced at high redshift.

\section{Galaxy stellar mass function (GSMF)}

\begin{figure}
	\includegraphics[width=\columnwidth]{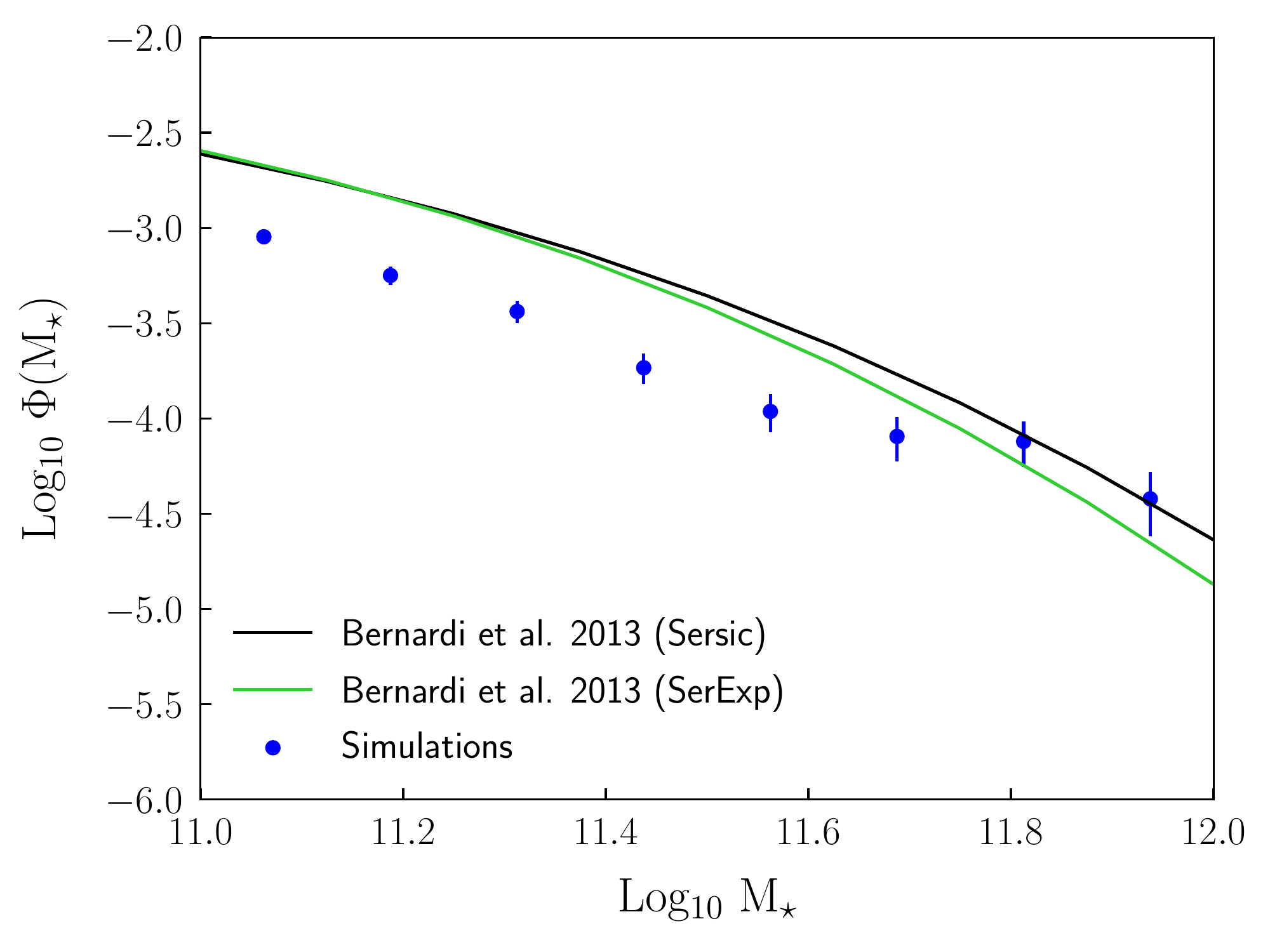}
    \caption{High mass end of Galaxy stellar mass function in simulations (blue points) compared to GSMF of \cite{2013MNRAS.436..697B}.}
    \label{fig17}
\end{figure}

We compare the high mass end of the GSMF in our simulations with respect to the results presented in \cite{2013MNRAS.436..697B}. In our simulations the GSMF is computed as follows: we considered the region of space within $R_{200}$ of all the 135 groups and clusters used in our paper. In this region we considered all the substructures but the BGGs and BCGs identified by Subfind and computed the stellar mass considering all the star particles which Subfind assign as bound to the substructure. The number of galaxies in each bin is than normalized following the prescription given in the Appendix of \cite{2014VULCANI} (equation B2). This factor takes into account the fact that we are in an overdense region with respect to the field. 

We note that simulations predict a GSMF lower by a factor of $\sim 2$ when comparing with observations at masses lower than $\sim 6 \times 10^{11}\ M_{\odot}$. This effect has been already discussed when studying the mismatch between the number of substructures in cluster environment as measured in observations and simulations (e.g., \citealt{2015ApJ...800...38G}; \citealt{2016ApJ...827L...5M}). 
\end{appendix}

\end{document}